\documentclass[pre, notitlepage, twocolumn, showpacs, floatfix, superscriptaddress]{revtex4-1}
\usepackage{amsmath}
\usepackage{amssymb}
\usepackage{amsthm}
\usepackage{amsfonts}
\usepackage{verbatim}
\usepackage{listings}
\usepackage{enumerate}
\usepackage{latexsym}
\usepackage{dsfont}
\usepackage{bm}
\usepackage[all]{xy}
\usepackage{graphicx}
\usepackage{color}
\usepackage{mathtools}
\usepackage[percent]{overpic}

\usepackage{epsfig,slashed}
\usepackage{epstopdf}
\usepackage{float}
\usepackage{mathtools}
\usepackage[colorlinks=true,citecolor=blue,linkcolor=blue,urlcolor=blue]{hyperref}
\usepackage{natbib}
\usepackage{subfigure}
\usepackage[mathscr]{euscript}
\usepackage{textcomp}
\usepackage{graphicx}
\usepackage{wasysym}
\usepackage{etoolbox}
\usepackage{lipsum}

\makeatletter
\appto\abstract{%
  \let\latexlist\list
  \def\list{\edef\keeprightskip{\the\rightskip}\latexlist}%
  \patchcmd\latexlist{\ignorespaces}{\rightskip\keeprightskip\ignorespaces}{}{}%
}

\makeatletter
\newsavebox\myboxA
\newsavebox\myboxB
\newlength\mylenA

\newcommand*\xoverline[2][0.75]{%
    \sbox{\myboxA}{$\m@th#2$}%
    \setbox\myboxB\null
    \ht\myboxB=\ht\myboxA%
    \dp\myboxB=\dp\myboxA%
    \wd\myboxB=#1\wd\myboxA
    \sbox\myboxB{$\m@th\overline{\copy\myboxB}$}
    \setlength\mylenA{\the\wd\myboxA}
    \addtolength\mylenA{-\the\wd\myboxB}%
    \ifdim\wd\myboxB<\wd\myboxA%
       \rlap{\hskip 0.5\mylenA\usebox\myboxB}{\usebox\myboxA}%
    \else
        \hskip -0.5\mylenA\rlap{\usebox\myboxA}{\hskip 0.5\mylenA\usebox\myboxB}%
    \fi}
\makeatother

\newcommand{\pdagger}{{\phantom{\dagger}}}

\newcommand{\mc}{\mathcal}

\renewcommand{\vec}[1]{\boldsymbol{#1}}

\newcommand{\equref}[1]{Eq.~(\ref{#1})}

\newcommand{\secref}[1]{Sec.~\ref{#1}}
\newcommand{\figref}[1]{Fig.~\ref{#1}}
\newcommand{\refcite}[1]{Ref.~\onlinecite{#1}}

\renewcommand{\approx}{\simeq}

\renewcommand{\vec}[1]{\boldsymbol{#1}}

\usepackage{braket}

\DeclareGraphicsExtensions{.png}
\begin{document}

\title{Electric-field-tunable electronic nematic order in twisted double-bilayer graphene}
\author{Rhine Samajdar}
\thanks{These two authors contributed equally.}
\affiliation{$\mbox{Department of Physics, Harvard University, Cambridge, MA 02138, USA}$}
\author{Mathias S. Scheurer}
\thanks{These two authors contributed equally.}
\affiliation{$\mbox{Institute for Theoretical Physics, University of Innsbruck, Innsbruck A-6020, Austria}$}
\author{Simon Turkel}
\affiliation{$\mbox{Department of Physics, Columbia University, New York, New York 10027, USA}$}
\author{Carmen Rubio-Verd\'{u}}
\affiliation{$\mbox{Department of Physics, Columbia University, New York, New York 10027, USA}$}
\author{Abhay N. Pasupathy}
\affiliation{$\mbox{Department of Physics, Columbia University, New York, New York 10027, USA}$}
\author{J\"{o}rn W. F. Venderbos}
\affiliation{$\mbox{Department of Physics, Drexel University, Philadelphia, PA 19104, USA}$}
\affiliation{$\mbox{Department of Materials Science and Engineering,
Drexel University, Philadelphia, PA 19104, USA}$}
\author{Rafael M. Fernandes}
\affiliation{$\mbox{School of Physics and Astronomy, University of Minnesota, Minneapolis, Minnesota 55455, USA}$}

\begin{abstract}
Graphene-based moir\'{e} systems have attracted considerable interest in recent years as they display a remarkable variety of correlated phenomena. Besides insulating and superconducting phases in the vicinity of integer fillings of the moir\'{e} unit cell, there is growing evidence for electronic nematic order both in twisted bilayer graphene and twisted double-bilayer graphene (tDBG), as signaled by the spontaneous breaking of the threefold rotational symmetry of the moir\'{e} superlattices. Here, we combine symmetry-based analysis with a microscopic continuum model to investigate the structure of the nematic phase of tDBG and its experimental manifestations. First, we perform a detailed comparison between the theoretically calculated local density of states and recent scanning tunneling microscopy data [arXiv:2009.11645] to resolve the internal structure of the nematic order parameter in terms of the layer, sublattice, spin, and valley degrees of freedom. We find strong evidence that the dominant contribution to the nematic order parameter comes from states at the moir\'{e} scale rather than at the microscopic scale of the individual graphene layers, which demonstrates the key role played by the moir\'e degrees of freedom and confirms the correlated nature of the nematic phase in tDBG. Secondly, our analysis reveals an unprecedented tunability of the orientation of the nematic director in tDBG by an externally applied electric field, allowing the director to rotate away from high-symmetry crystalline directions. We compute the expected fingerprints of this rotation in both STM and transport experiments, providing feasible ways to probe it.
Rooted in the strong sensitivity of the flat bands of tDBG to the displacement field, this effect opens an interesting route to the electrostatic control of electronic nematicity in moir\'{e} systems.
\end{abstract}
\maketitle

\hypersetup{linkcolor=blue}


\section{Introduction}

Moir\'e superlattice materials have rapidly emerged as versatile platforms for studying strong correlation effects in highly tunable two-dimensional electronic systems \cite{macdonald2019bilayer, andrei2020graphene,kennes2020moir}. Their low-energy electronic structure is determined by, and very sensitive to, the superlattice moir\'e potential originating from the rotational misalignment of the constituent layers and/or the mismatch between the lattice constants of these layers. Common to many known moir\'e superlattice systems---such as twisted bilayer graphene (tBG) \cite{2018Natur.556...80C,SuperconductivityTBG,Yankowitz1059,2019arXiv190306513L}, $\mathrm{ABC}$-trilayer graphene with hexagonal boron nitride substrate (TLG/hBN) \cite{SuperconductivityInTrilayer,chen2020tunable}, twisted double-bilayer graphene (tDBG) \cite{2019arXiv190306952S,ExperimentKim,PabllosExperiment,burg2019correlated}, or transition metal dichalcogenide (TMD) heterostructures \cite{zhang2020flat,wang2020correlated}---is the presence of isolated moir\'e minibands with dramatically reduced bandwidths, often referred to as flat bands \cite{balents2020superconductivity}. Since interaction effects are comparatively strong in these narrow bands, one generally expects the emergence in moir\'e systems of phenomena  typically observed in strongly correlated bulk materials. 

The most prominent correlation-driven phase observed in different multilayer moir\'e systems is the correlated insulating state found at integer fillings of the moir\'e unit cell \cite{2018Natur.556...80C,codecido2019correlated}. Additionally, tBG \cite{SuperconductivityTBG,Yankowitz1059,2019arXiv190306513L}, TLG/hBN \cite{SuperconductivityInTrilayer}, and very recently, twisted mirror-symmetric trilayer graphene \cite{MirrSymTrilayer1,MirrSymTrilayer2}, also show robust evidence for superconductivity \cite{ chu2020superconductivity}, inspiring several theoretical studies \cite{fidrysiak2018unconventional, PhysRevB.98.195101, 2018PhRvB..98v0504P,2018PhRvL.121y7001W, 2018PhRvB..98x1412C,RafaelsPaperTBG,2018PhRvX...8d1041I,2018PhRvB..98g5154D,2018PhRvB..98t5151S, lin2018kohn,CenkeLeon,2018PhRvL.121u7001L,2018PhRvB..98x1407K, AshvinModelTBG,2018PhRvB..98g5109T, lian2019twisted, alidoust2019symmetry, wu2019topological, YiZhuangPairing, hu2019geometric, huang2019antiferromagnetically,julku2019superfluid,PhysRevB.99.195120,2019PhRvB..99m4515R, 2019arXiv190903514W,PhysRevLett.122.026801,PhysRevB.100.085136,PhysRevB.101.155413,PhysRevB.103.L041103,dai2020mott,2019arXiv190603258S,OurMicroscopicTDBG,PhysRevB.103.024506}. On the other hand, manifestations of superconducting behavior in other moir\'e systems, such as tDBG \cite{2019arXiv190603258S,OurMicroscopicTDBG,2019arXiv190308685L, 2019arXiv190607302W,li2019phonon,hsu2020topological}, appear to be more fragile \cite{PabllosExperiment,burg2019correlated}. More recently, evidence for a universal rotational-symmetry-breaking phase at small twist angles has also been reported in tBG \cite{PasupathySTM,AndreiSTM,NadjPergeSTM,STMReview,Cao2020_nematics}, tDBG \cite{rubioverdu2020universal}, and TMD heterostructures \cite{Jin2020_stripe}. The broken symmetry, which seems to occur only at specific---but not necessarily commensurate---carrier concentrations, is the characteristic threefold rotational symmetry of the triangular moir\'e superlattice. 

Detailed scanning tunneling microscopy (STM) and spectroscopy (STS) data \cite{PasupathySTM,AndreiSTM,NadjPergeSTM,STMReview}, as well as transport measurements \cite{Cao2020_nematics}, provide strong evidence for an intrinsic electronic nematic instability rather than an extrinsic origin for this effect, such as heterostrain \cite{PhysRevB.100.035448,PhysRevB.100.205113,PhysRevResearch.2.023325,uri2020mapping}. Importantly, due to the hexagonal or trigonal symmetry of the graphene and TMD layers that make up the moir\'e superlattice, such nematic order is not of the conventional Ising type observed in bulk tetragonal correlated systems \cite{fradkin2010nematic,NematicityReview,PhysRevB.93.064520,PhysRevLett.118.227601}. Instead, it is described by a two-component nematic order parameter \cite{2019arXiv191111367F}, making it a much closer analogue of the classical nematic liquid-crystalline phase. While the observation of threefold rotational-symmetry breaking in three different classes of moir\'e systems hints at a possible universal character of the electronic nematic phase, many questions remain open, such as the microscopic mechanism behind this instability and its relevance for the insulating and superconducting states observed in the phase diagrams. Theoretically, much of the focus has been on nematicity in tBG, both in the normal state \cite{RafaelsPaperTBG,2019arXiv191111367F,Vishwanath2021,Brillaux2020, PhysRevB.102.205111,PhysRevB.102.035161,2018PhRvB..98g5154D,2018PhRvX...8d1041I,Bascones2020,Sboychakov2020,Chichinadze_nematics,Bernevig_TBG_VI} and in the superconducting state \cite{PhysRevB.98.195101, 2019arXiv190603258S,2019arXiv190603258S,LiangFu,2019arXiv191007379C,PhysRevB.103.024506,yu2021nematicity}, while the specific nematic properties of tDBG and TMD heterostructures remain little explored. 

In this work, motivated by recent STM and STS measurements in tDBG \cite{rubioverdu2020universal}, we investigate the microscopic properties and macroscopic manifestations of electronic nematicity in this particular moir\'e system. tDBG consists of two Bernal-stacked (i.e.,~AB-stacked) graphene bilayers that are twisted by an angle $\theta$ relative to one another [see \figref{fig:Lattice}(a)]. Phenomenologically, breaking of the threefold rotational symmetry ($C_3$) of tDBG can be accomplished by a large number of combinations of the available degrees of freedom: layer, sublattice, spin, and valley. Moreover, the $C_3$ symmetry can be broken at either the atomic scale of the graphene layers (which we dub graphene nematicity) or at the emergent scale of the moir\'e superlattice (which we refer to as moir\'e nematicity). Distinguishing between these two scenarios is important to elucidate whether the nematic instability observed in tDBG is inherited from an underlying instability of bilayer graphene or whether it is an actual correlation-driven instability associated with the flat bands. Indeed, it has been previously shown that bilayer graphene on its own can undergo a nematic transition \cite{Mayorov2011,PhysRevB.86.075467}.

To answer this question, we adopt the frequently used \cite{OurMicroscopicTDBG,rubioverdu2020universal,zhang2020visualizing,liu2020spectroscopy,koshino2019band} continuum-model \cite{koshino2019band,dos2007graphene,bistritzer2011moire,dos2012continuum} description of tDBG and identify the different
possible nematic channels (i.e., irreducible representations with broken $C_3$ symmetry but preserved time-reversal and translational symmetries). Within each channel, we examine the most
natural microscopic forms of the nematic order parameter
consistent with the symmetry of the nematic channel; in particular, we focus on moir\'e nematicity, intravalley
graphene nematicity, and intervalley graphene nematicity.
While terms in the same channel transform under the same representation and hence, symmetry alone cannot distinguish between them, it is still sensible and relevant to ask whether there is a dominant order parameter.

We address this issue via a detailed comparison between the energy and position dependence of the STM $\mathrm{d}I/\mathrm{d}V$ maps reported in \refcite{rubioverdu2020universal} with that of the theoretical local density of states (LDOS) characterizing the nematic phase generated by each nematic order parameter considered here---namely, moir\'e nematicity, intravalley graphene nematicity, and intervalley graphene nematicity. We show that the main quantitative and qualitative features observed in the STM data (such as the fact that the nematic anisotropy is maximum at energies corresponding to the valence flat band) can be very well described by a simple moir\'e nematic order parameter, and are not captured by the other graphene nematic order parameters. While there are many other possible forms for the order parameter that are not discussed here, the strong agreement between the theoretical LDOS and the experimental data suggests that the nematic instability in tDBG is a property of the emergent moir\'e degrees of freedom, and is thus likely to be on an equal footing with the correlated insulating state observed in this system. We note that while some of these conclusions have been put forward by us in \refcite{rubioverdu2020universal}, the present work considerably expands the theoretical analysis performed in that work.

A unique characteristic of tDBG, especially when compared to tBG, is the high degree of tunability of its electronic band structure by a perpendicular displacement field \cite{2019arXiv190108420R,2019arXiv190300852C,2019PhRvB..99g5127Z,koshino2019band,FirstModel,2019arXiv190600623H}. 
Our symmetry analysis reveals that the electric field couples to the nematic director of a generic two-component nematic order parameter via a linear-cubic coupling in the free-energy expansion, thereby allowing the director to rotate away from the three high-symmetry crystalline directions. Focusing on the case of moir\'e nematicity, as supported by the STM data, we microscopically compute the coupling constant for tDBG and show that the director can rotate over a substantial range of about $10^{\circ}$ for experimentally accessible values of the displacement field. We further discuss the predicted experimental manifestations of the nematic director's rotation in STM \cite{rubioverdu2020universal,zhang2020visualizing,liu2020spectroscopy} and transport \cite{PhysRevB.101.125428,PhysRevLett.125.176801,EnsslinExpDW} measurements, disentangling it from effects caused by the changes in the band structure prompted by a changing displacement field. These results demonstrate the unique capabilities of moir\'e systems for electrostatic control of electronic nematicity.

This paper is organized as follows: we begin by carefully addressing the microscopic origin of the observed nematic order in tDBG by examining different possible microscopic realizations of nematic order (Sec.~\ref{AFewNaturalMicroscopicForms}). We then proceed by demonstrating that anisotropic charge redistribution on the moir\'e scale is consistent with---and best describes---the experimental data, as opposed to charge redistribution on the scale of the constituent graphene layers (Sec.~\ref{sec:LDOS}). In the final part of the paper, we turn to the electric-field control of nematic order (Sec.~\ref{sec:EField}). Finally, concluding remarks are discussed in Sec.~\ref{sec:conclusions}. For completeness, all details of the continuum model used in this work are summarized in Appendix~\ref{app:model}.

\section{Microscopic structure of the nematic order parameter}

\label{AFewNaturalMicroscopicForms}

\begin{figure}[tb]
\includegraphics[width=\linewidth]{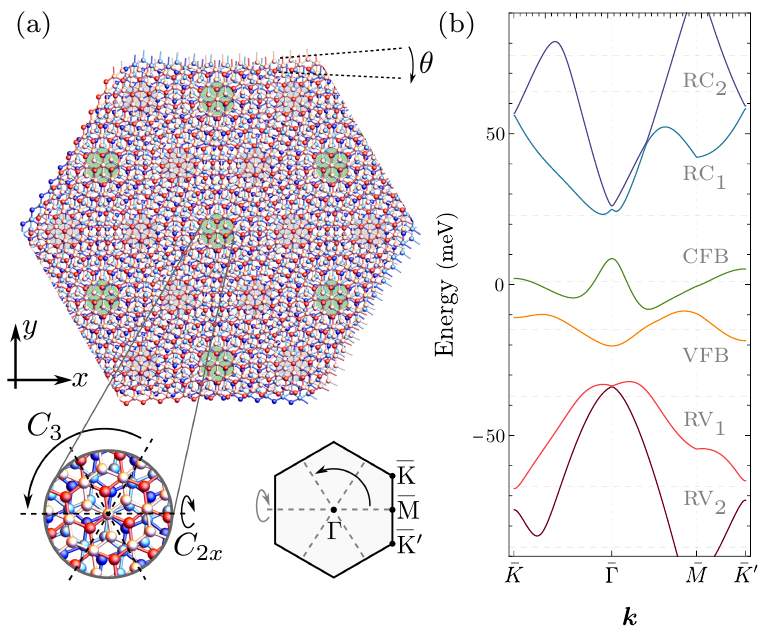} 
\caption{\label{fig:Lattice}(a) The four-layered tDBG heterostructure is created by twisting two Bernal-stacked graphene bilayers by a relative angle $\theta$, thus forming a moir\'{e} superlattice. The latter is invariant under threefold rotations ($C_3$) along the $z$ axis and twofold rotations ($C_{2x}$) with respect to the $x$ axis (left inset). The moir\'e mini-Brillouin zone is sketched in the right inset. (b) Low-energy band structure of tDBG obtained from the continuum model for a twist angle of $\theta$\,$=$\,$1.05^\circ$ using the parameters described in \refcite{rubioverdu2020universal}, and reproduced in Appendix~\ref{app:model}. The valence (conduction) flat band is denoted by VFB (CFB) while RV$_j$ (RC$_j$) labels the remote valence (conduction) bands. Note that the applied displacement field is responsible for the energy gap between the flat bands at the charge neutrality point. The Fermi energy is chosen so as to correspond to a filling fraction of the CFB of $\nu$\,$=$\,$0.475$, for which nematic order was experimentally observed to be the strongest \cite{rubioverdu2020universal}. The dashed lines mark the energies analyzed in Figs.~\ref{fig:Maps} and \ref{fig:LDOS}. }
\end{figure}

We first employ symmetry considerations to discuss the possible microscopic structures of the nematic order parameter in terms of the electronic degrees of freedom present in tDBG. Generically, an electronic nematic phase is defined by a correlation-driven lowering of the point group symmetry of a crystal while preserving translational symmetry \cite{fradkin2010nematic,NematicityReview,fernandes2019intertwined}. The structure of tDBG has point group $D_3$: as illustrated in Fig.~\ref{fig:Lattice}(a), it is invariant under threefold rotations perpendicular to the graphene sheets, $C_3$, as well as a twofold in-plane rotation along the $x$ axis, $C_{2x}$. However, the latter symmetry is broken when a vertical displacement field $D$ is applied along the $z$ axis. This electric field has a strong effect on the electronic band structure, affecting both the energy gaps between the bands and the bandwidths \cite{2019PhRvB..99g5127Z,2019arXiv190108420R,2019arXiv190300852C,koshino2019band,FirstModel,2019arXiv190600623H}. As a result, the displacement field is used in experiments as an integral tuning parameter for controlling the moir\'{e} flat bands \cite{2019arXiv190306952S,ExperimentKim,PabllosExperiment,burg2019correlated,PhysRevB.101.125428}. As $D$ transforms under the $A_2$ representation of $D_3$, a finite displacement field lowers $D_3$ down to the point group $C_3$.

As opposed to tBG, the geometry of the carbon lattices in tDBG shows that twofold rotation perpendicular to the graphene layers, $C_2$, is not an approximate symmetry at small twist angles $\theta$ and is, in fact, strongly broken by the lattice. Accordingly, the combined symmetry $C_2\Theta$, involving time-reversal, $\Theta$, is not present either. This explains the absence of Dirac cones in the band structure of tDBG, which is shown in Fig.~\ref{fig:Lattice}(b). Here, we used the experimental parameters for which nematic order was found to be the strongest \cite{rubioverdu2020universal}. In particular, the chemical potential is set to slightly below half-filling ($\nu=0.475$) of the conduction flat band (CFB), which is energetically close to but separated from the valence flat band (VFB). Figure~\ref{fig:Lattice}(b) also displays two occupied and two unoccupied remote bands (RV$_j$ and RC$_j$, respectively), and focuses on only one valley, since the bands arising from the other valley can be obtained by transformation under $\Theta$. Note that the absence of $C_2$ symmetry in tDBG has further-reaching consequences, e.g., for the possible superconducting phases and the pairing mechanisms \cite{2019arXiv190603258S,2020arXiv200400638K,Christos29543}.

For the purpose of our symmetry analysis in tDBG, we start from the limit without a displacement field ($D$\,$=$\,$0$), where the system's point group is $D_3$.
The nematic order parameter is then a vector $\vec{\Phi}$\,$=$\,$(\Phi_1,\Phi_2)^T$\,$\in$\,$\mathbb{R}^2$ that transforms under the two-dimensional $E$ irreducible representation (``irrep'') of $D_3$ and is time-reversal even. 
Such a vector is naturally represented as $\vec{\Phi}$\,$=$\,$\Phi\,(\cos 2\varphi, \sin 2\varphi)$, where the angle $\varphi$ can be identified with the orientation of the nematic director $\hat{n}$\,$=$\,$(\cos \varphi, \sin \varphi)$. Note that $\varphi$ and $\varphi +\pi$ both result in the same nematic order parameter, as expected. A nonzero $\Phi$ breaks the threefold rotational symmetry, $C_3$, of the moir\'e superlattice.
Choosing, without loss of generality, $\Phi_1$ and $\Phi_2$ to transform as $x$ and $y$ under $D_3$, and defining the complex-valued order parameter $\hat{\Phi}$\,$\equiv$\,$\Phi_1 + i\, \Phi_2$, one finds the transformation properties:
\begin{subequations}
\begin{align}
    C^{}_3:\quad \hat{\Phi} \, &\longrightarrow \, \omega \,\hat{\Phi}, \quad \omega \equiv e^{2i\pi/3}, \\
    C^{}_{2x}: \quad \hat{\Phi} \, &\longrightarrow \, \hat{\Phi}^*.
    \label{eq:C2x}
\end{align} \label{ActionOfSyms}
\end{subequations}
As discussed earlier, the twofold in-plane rotational symmetry described by Eq.~\eqref{eq:C2x} is broken by a finite displacement field, which---as we will show in Sec.~\ref{sec:EField}---allows one to rotate the nematic director away from the high-symmetry directions.
For now, we assume that the $C_{2x}$ symmetry is already broken explicitly by the electric field and study the properties of the LDOS inside the nematic phase as a function of the \textit{a priori} arbitrary orientation of the nematic director.

We begin our discussion of the possible microscopic nematic order parameters via a real-space description using the continuum model \cite{koshino2019band} (additional details can be found in Appendix~\ref{app:model}). Importantly, this model has been shown to reproduce well many experimental features of the band structure of tDBG \cite{ExperimentKim,PabllosExperiment,rubioverdu2020universal,zhang2020visualizing,liu2020spectroscopy}. In order to make the comparison with the STM data more accurate, in this section we consider a self-consistently calculated screened electric field \cite{rubioverdu2020universal}, instead of assuming a uniform difference in electrostatic energy between adjacent layers. In the continuum model, the relevant electronic field operators $c_{\sigma, \ell, s, \eta}(\vec{r})$ refer to electrons on the graphene layers that constitute the tDBG device, and are thus characterized by spin $\sigma = \uparrow, \downarrow$, sublattice $s$\,$=$\,$A, B$, layer $\ell $\,$=$\,$1,2,3,4$, valley $\eta$\,$=$\,$\pm$, and (continuum) position $\vec{r}\in\mathbb{R}^2$. Within this framework, the nematic order parameter $\vec{\Phi}$ will couple to two fermionic bilinears in the particle-hole channel (of the form $c^\dagger c$) that transform as the two components of the two-dimensional $E$ irrep of $D_3$. Since spin-orbit coupling (SOC) is small in graphene and we do not expect the onset of nematic order in tDBG to be accompanied by the simultaneous breaking of spin-rotational symmetry, we focus on spin-unpolarized nematic order parameters. Taken together, the coupling must be of the form
\begin{align}\begin{split}
        \mathcal{H}_{\vec{\Phi}} = \int_{\vec{r}} &\int_{\Delta\vec{r}}\,\, \vec{\Phi}\cdot  \vec{\phi}^\pdagger_{\ell,s,\eta;\ell',s',\eta'}(\vec{r},\Delta\vec{r}) \\
        &\quad\times c^\dagger_{\sigma, \ell, s, \eta}(\vec{r}+\Delta\vec{r}) c^\pdagger_{\sigma ,\ell',s',\eta'}(\vec{r}) + \text{H.c.}\,;\label{GeneralFormOfNematicOrder}
\end{split}\end{align}
note that this form of $\mathcal{H}_{\vec{\Phi}}$ is completely general and does not rely on any additional microscopic details.
The two-component, matrix-valued function of two spatial coordinates $\vec{\phi}_{s,\ell,\eta;s',\ell',\eta'}(\vec{r},\Delta\vec{r})$ is only constrained by $C_3$ (as well as $C_{2x}$ if $D=0$), time-reversal symmetry, and moir\'e translations $\vec{\phi}(\vec{r}+\vec{a}^{\textsc{m}}_{j},\Delta\vec{r}) = \vec{\phi}(\vec{r},\Delta\vec{r})$. Here, $\vec{a}^{\textsc{m}}_{j}\propto (\sqrt{3},-(-1)^j)^{\textsc{T}}$, $j$\,$=$\,$1,2$, are the primitive vectors of the moir\'e superlattice. 

The microscopic structure of the nematic order is encoded in the complicated nematic coupling function $\vec{\phi}_{s,\ell,\eta;s',\ell',\eta'}(\vec{r},\Delta\vec{r})$, which must transform as the $E$ irrep, like $\vec{\Phi}$. Crucially, the $C_3$ symmetry operation acts not only on the internal degrees of freedom $s,\ell,\eta$ but also on the real-space structure of this function. Our goal will be to learn about the form of $\vec{\phi}$ from the experimental STM data of \refcite{rubioverdu2020universal}. This will yield information about not only the microscopic structure of the nematic order but also the type of interactions (e.g., on the microscopic graphene scale of the individual bilayers or on the scale of the emergent moir\'e lattice) responsible for it.
However, due to the multitude of internal degrees of freedom available---layer, sublattice, and valley---and the (technically infinite) variety of possible combinations of basis functions that transform as $E$, it is challenging to extract the exact form of $\vec{\phi}$ directly from experiments. Thus, to shed light on the structure that best describes the STM data, 
we will consider three simple \emph{ans\"atze} for $\vec{\phi}$.
Specifically, we concentrate on two opposite limits. The nematic order parameter can break $C_3$ symmetry at the scale of either the moir\'{e} superlattice (e.g., by making the effective tunneling of flat-band electrons between moir\'e unit cells directionally inequivalent) or the underlying carbon lattice of graphene. In the former case, it is the spatial dependence of $\vec{\phi}$ that transforms nontrivially under $C_3$, whereas in the latter case, it is its dependence on the internal degrees of freedom. We refer to these two options as ``moir\'{e} nematic'' and ``graphene nematic'', respectively. As the name suggests, moir\'{e} nematicity is insensitive to the local properties of the graphene layers. On the other hand, graphene nematicity focuses exclusively on the atomic scale, and is affected by the existence of the moir\'{e} superlattice only due to the imposed interlayer potential. Below, we will further distinguish two different types of graphene nematicity.

\subsection{Moir\'{e} nematic order}

Although we work with a continuum model, it is natural to consider the possibility that the nematicity itself occurs at the moir\'e scale and hence, stems from an anisotropic deformation of the effective hopping amplitudes between different moir\'e unit cells. To capture this possibility, we focus on $\vec{\phi}\,(\vec{r},\Delta \vec{r})$ in \equref{GeneralFormOfNematicOrder} with $\Delta \vec{r}$ belonging to the moir\'e superlattice, i.e., $\Delta \vec{r} = \vec{R}_{p,q}$\,$=$\,$ p\,\vec{a}^{\textsc{m}}_{1}$\,$+$\,$q\,\vec{a}^{\textsc{m}}_{2}$ ($p, q$\,$\in$\,$\mathbb{Z}$). As the $C_3$-symmetry breaking is encoded in the spatial dependence of $\vec{\phi}$, we can restrict the nematic coupling function to be trivial (i.e., diagonal) in all other internal indices $\alpha\equiv (\sigma, \ell, s, \eta)$. Equation~(\ref{GeneralFormOfNematicOrder}) then becomes
\begin{equation}
        \mathcal{H}^{\textsc{m}}_{\vec{\Phi}} = \frac{1}{2} \int_{\vec{r}} \sum_{p,q\,\in\,\mathbb{Z}}\vec{\Phi}\cdot  \vec{\phi}^{}_{p,q}(\vec{r})\, c^\dagger_{\alpha}(\vec{r}+\vec{R}_{p,q}) \,c^\pdagger_{\alpha}(\vec{r}) + \text{H.c.}\,. \label{HPhiCoupl}
\end{equation}

\begin{figure}[b]
\includegraphics[width=\linewidth]{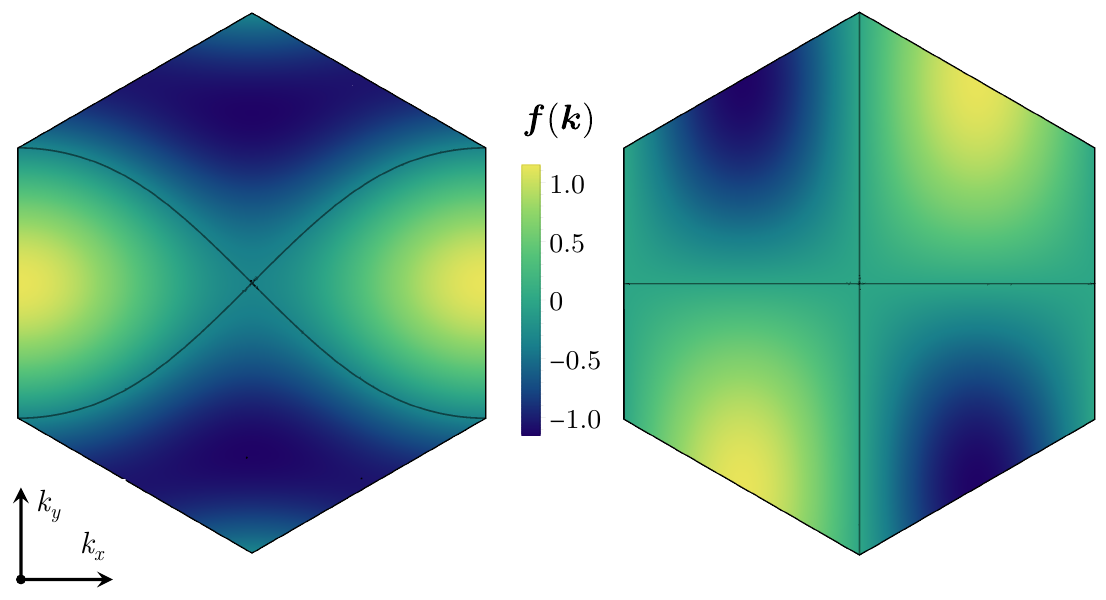}
\caption{\label{fig:basis}Momentum dependence of the two components of the function $\vec{f}(\vec{k})$. The first (second) component, as defined in Eq.~\eqref{LowestOrderBasisFuncs}, is plotted in the left (right) panel. The black curves correspond to the nodal lines across which the functions change sign, as can also be seen from the color map.}
\end{figure}

Time-reversal and translational symmetries enforce, respectively, that $\vec{\phi}_{p,q}(\vec{r}) \in \mathbb{R}^2$ and that $\vec{\phi}_{p,q}(\vec{r} + \vec{a}^{\textsc{m}}_{j}) = \vec{\phi}_{p,q}(\vec{r})$. In the absence of any prior knowledge about the detailed functional form of $\vec{\phi}_{p,q}(\vec{r})$ on the continuum coordinate $\vec{r}$, we work with an $\vec{r}$-independent basis function such that $\vec{\phi}_{p,q}(\vec{r})$\,$=$\,$\vec{\phi}_{p,q}$. Thus, it is the dependence of $\vec{\phi}_{p,q}$ on $p$ and $q$ that makes it transform under the $E$ irrep of $D_3$. With this choice of a spatially uniform $\vec{\phi}_{p,q}$, the nematic coupling in Eq.~\eqref{HPhiCoupl} simply leads to a momentum-dependent shift of the energy spectrum $E_{n,\vec{k}}$ of the continuum model
\begin{equation}
    E_{n,\vec{k}} \, \rightarrow \, E_{n,\vec{k}} + \vec{\Phi}\cdot\vec{f}(\vec{k}), \,\, \vec{f}(\vec{k}) = \sum_{p,q\in\mathbb{Z}} \vec{\phi}^{}_{p,q} \cos (\vec{k} \cdot \vec{R}^{}_{p,q}). \label{ShiftOfTheBandstructureByOP}
\end{equation}
Here, $\vec{k}$ is the moir\'e momentum, which can be chosen to be in the first Brillouin zone (FBZ) of the moir\'e superlattice, and $n$ labels the (technically infinitely many) bands of the continuum model without nematic order.
Note that the combination of time-reversal symmetry and our restriction, in Eq.~\eqref{HPhiCoupl}, to only terms that are diagonal in $\alpha$ results in $\vec{f}(\vec{k})$\,$=$\,$\vec{f}(-\vec{k})$.

Constraining \equref{HPhiCoupl} to terms involving hopping matrix elements between only neighboring moir\'e unit cells, and enforcing $\vec{\phi}_{p,q}$ to transform as the $E$ irrep, we obtain
\begin{eqnarray}
& & \vec{\phi}_{p,q} =
\label{NearestNeighbor} \\
& & \frac{8}{3}\bigg[\delta_{p,1}\delta_{q,-1} - \delta_{p,1}\delta_{q,0} - \delta_{p,0}\delta_{q,1},\frac{\sqrt{3}}{2}\left(\delta_{p,0}\delta_{q,1}-\delta_{p,1}\delta_{q,0}\right) \hspace*{-0.1cm}\bigg]^\mathrm{T}. \nonumber
\end{eqnarray}
Thereupon, setting $\vert\vec{a}_{M,r}\vert$\,$=$\,$1$, we find
\begin{eqnarray}
& & \vec{f}(\vec{k}) =
\label{LowestOrderBasisFuncs} \\
& & \frac{8}{3}\left( \cos k_y - \cos \frac{\sqrt{3}k_x}{2} \cos \frac{k_y}{2} , \sqrt{3} \sin \frac{\sqrt{3}k_x}{2} \sin \frac{k_y}{2}  \right)^\mathrm{T}. \nonumber
\end{eqnarray}
The momentum dependence of the two components of $\vec{f}(\vec{k})$ in the FBZ is illustrated in Fig.~\ref{fig:basis}. As expected for a function that transforms as the $E$ irrep, it behaves as $\vec{f}(\vec{k}) \sim (k_x^2-k_y^2,2k_xk_y)^\mathrm{T}$ for small $\vec{k}$.

\begin{figure}[tb]
\includegraphics[width=0.95\linewidth]{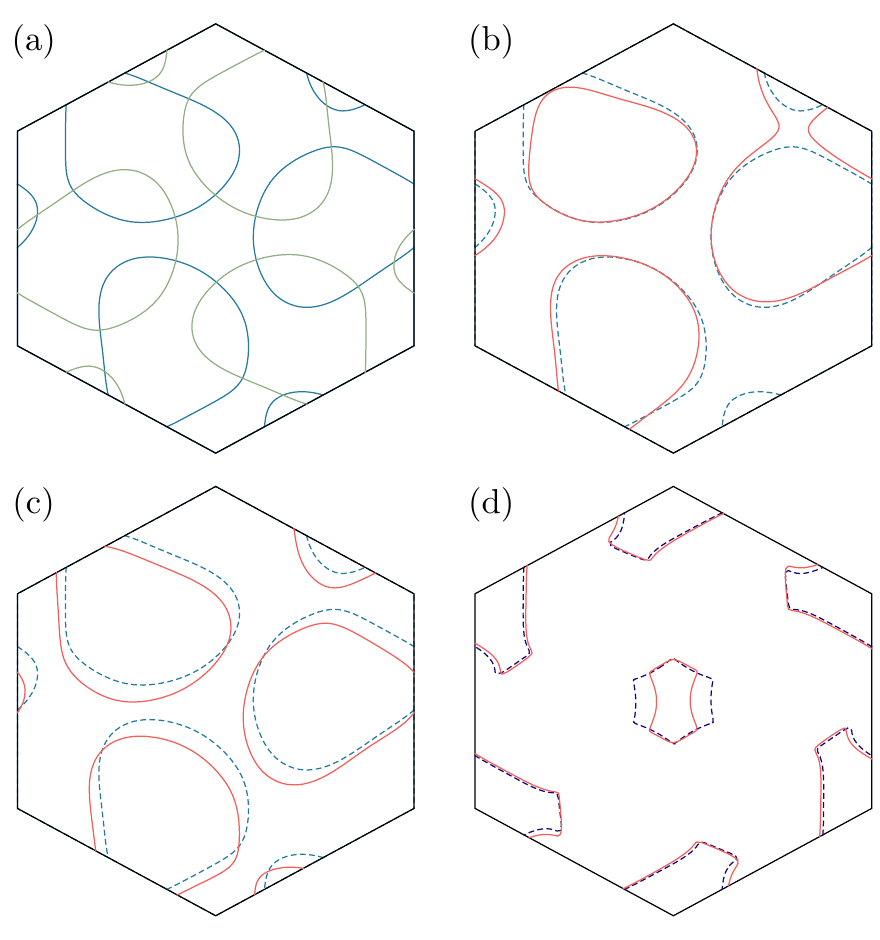}
\caption{\label{fig:FS}(a) Fermi surfaces in the absence of nematic order at a filling of $\nu$\,$=$\,$0.475$, corresponding to a chemical potential that crosses the conduction flat band of Fig.~\ref{fig:Lattice}(b). The Fermi surfaces are obtained by diagonalizing the Hamiltonian independently in the two valleys, $\eta$\,$=$\,$+$ (blue) and $\eta$\,$=$\,$-$ (green), and are symmetric under $C_3$. (b, c) Focusing on the $\eta$\,$=$\,$+$ valley's Fermi surfaces (red lines), the $C_3$ symmetry of the original Fermi surfaces (blue dashed lines) is broken, albeit in different manners, by the (b) moir\'{e} nematic ($\Phi$\,$=$\,$0.5$\,meV) and (c) intravalley graphene nematic ($\Phi$\,$=$\,$20$\,meV) order. (d) In the presence of intervalley graphene nematic order, the valleys are no longer decoupled and the Fermi surfaces of both valleys must be considered together. Even without nematicity (blue dashed lines), the Fermi surfaces of panel (a) are reconstructed due to the intervalley coupling. In this case, the main effect of nematic order (set here to be of strength $\Phi$\,$=$\,$50$\,meV) is seen near the $\Gamma$ point (red lines).}
\end{figure}

Figure~\ref{fig:FS}(a) illustrates the Fermi surface in the absence of nematic order for a fractional occupation ($\nu$\,$=$\,$0.475$) of the CFB such that the chemical potential crosses the green flat band in Fig.~\ref{fig:Lattice}(b). The moir\'e unit cell can be occupied by up to $n_s$\,$=$\,$4$ electrons per band, associated with the two possible values of the spin and valley quantum numbers. We use the convention that $\nu$\,$=$\,$0$ at charge neutrality whereas $\nu$\,$=$\,$\pm 1$ corresponds to full/empty flat bands. In Fig.~\ref{fig:FS}(a), the blue and green Fermi surfaces correspond to the two different valleys, and are therefore related by a twofold rotation. Since the continuum model possesses U$(1)_v$ valley symmetry, Fermi surfaces stemming from different valleys can cross. In Fig.~\ref{fig:FS}(b), for simplicity, we show only the Fermi surfaces corresponding to one of the valleys, both in the absence (blue dashed lines) and presence (red solid lines) of moir\'e nematic order. The breaking of $C_3$ symmetry is evident and manifested as distortions of the Fermi pockets.

The fact that the moir\'e nematic order parameter corresponds to a simple momentum-dependent shift of the energy dispersion that is the same for all bands $n$ in Eq.~\eqref{ShiftOfTheBandstructureByOP} does not imply that its effect on physical quantities, such as the STM $\mathrm{d}I/\mathrm{d}V$ maps, is the same for all bands. In general, we expect that its effect is the strongest at energy levels either close to a van-Hove singularity or whose Bloch wavefunctions change rapidly as a function of momentum (which generically occurs when two bands come close to each other).

\subsection{Intravalley graphene nematic order}
   
The moir\'e nematic order parameter has a characteristic momentum dependence, but is trivial in all internal degrees of freedom (layer, valley, spin, and sublattice). We now consider the opposite limit of an entirely local order parameter, $\vec{\phi}(\vec{r},\Delta\vec{r}) \propto \delta(\Delta\vec{r})$ in \equref{GeneralFormOfNematicOrder}, whose nontrivial transformation properties under $C_3$ follow only from a nontrivial structure in the internal degrees of freedom. We name this type of order parameter graphene nematicity to emphasize its local character.

We start by noting that the U(1)$_v$ valley transformation acts on the electronic operators as
\begin{equation}
    \text{U(1)}_v:\quad c\,(\vec{r}) \, \rightarrow \, e^{i\eta_z \vartheta }c\,(\vec{r}), \label{U1vSymmetry}
\end{equation}
where $\eta_j$ are Pauli matrices in valley space and $\vartheta$ is an arbitrary rotation angle. The nematic order parameter must transform under an irrep of $\text{U(1)}_v$ and, hence, can be separated into valley-diagonal, $\propto \eta_{0,z}$, [trivial representation of U(1)$_v$] or intervalley coherent nematic order, i.e., order parameters involving $\eta_{x,y}$ [two-dimensional representation of U(1)$_v$]. To proceed, we discuss these two cases separately, and first examine the case of intravalley (i.e., valley-diagonal) order, deferring the discussion of intervalley order to Sec.~\ref{sec:IVC} below.

In order to construct the nematic order parameter, we further need to know the representation of the relevant symmetry operations on the internal degrees of freedom. Assuming $C_{2x}$ is broken explicitly by the displacement field, these are time-reversal ($\Theta$) and threefold rotational  ($C_3$) symmetries. Using the conventions of Refs.~\cite{OurMicroscopicTDBG,koshino2019band}, we have 
\begin{equation}
\Theta\, c\,(\vec{r})\, \Theta^\dagger = i\sigma^{}_y\eta^{}_x c\,(\vec{r}) \label{TimeReversalRepr}
\end{equation}
for the action of the anti-unitary time-reversal symmetry and
\begin{equation}
    C_3:\quad c\,(\vec{r}) \, \rightarrow \, e^{i\frac{2\pi}{3}(\rho_z-\mc{L})\eta_z} c\,(C_3\vec{r}) \label{C3Repr}
\end{equation}
for the (unitary) threefold rotational symmetry. Here, $\sigma_j$ and $\rho_j$ represent Pauli matrices in spin and sublattice spaces, respectively. Moreover, $\mc{L}$ is defined as  $\mc{L}=\sigma_0\,\eta_0\, \mathrm{diag}(1,0,0,-1)$, where ``$\mathrm{diag}$'' stands for a diagonal matrix with the four indicated entries, which are associated with the four layers $\ell=1,2,3,4$. It is straightforward to check that the continuum Hamiltonian defined in \refcite{OurMicroscopicTDBG} and used in our numerics is invariant under Eqs.~\eqref{TimeReversalRepr} and \eqref{C3Repr}.

We are now in position to construct graphene (i.e., local) nematic order parameters that transform as the $E$ irrep. For the case of intravalley order parameters, \equref{GeneralFormOfNematicOrder} assumes the form
\begin{equation}
        \mathcal{H}^{\textsc{g}}_{\vec{\Phi}} = \int_{\vec{r}}\, \vec{\Phi}\cdot  \vec{\phi}^\pdagger_{\ell, s;\ell', s'}(\eta;\vec{r}) \, c^\dagger_{\sigma,\ell, s, \eta}(\vec{r}) c^\pdagger_{\sigma,\ell', s', \eta}(\vec{r}), \label{HPhiCoupl2}
\end{equation}
where $\vec{\phi}^*_{\ell', s';\ell, s}(\eta;\vec{r})$\,$=$\,$\vec{\phi}_{\ell, s;\ell', s'}(\eta;\vec{r})$ and $\eta$\,$=$\,$ \pm$ is the valley quantum number. Focusing on combinations that are also diagonal in the layer index $\ell$, i.e., devoid of any layer-hopping component, and taking, as above, $\vec{\phi}_{\ell, s;\ell', s'}(\eta;\vec{r})$ in Eq.~\eqref{HPhiCoupl2} to be independent of $\vec{r}$, we arrive at 
\begin{equation}
\vec{\phi}^\pdagger_{\ell, s;\ell', s'}(\eta) = \delta^\pdagger_{\ell,\ell'}\psi^\pdagger_{\ell}  \begin{pmatrix}(e^{i\alpha_\ell \eta \rho_z} \rho^\pdagger_x)^\pdagger_{ss'}\\ \eta(e^{i\alpha_\ell \eta \rho_z}\rho^\pdagger_y)^\pdagger_{ss'}\end{pmatrix}
\label{OnsiteOrderParameter}
\end{equation}
as the only remaining combination that displays the correct behavior under $C_3$ and $\Theta$.
Here, $\psi_{\ell}, \alpha_\ell \in\mathbb{R}$ are undetermined layer weights and layer-rotation angles. Pursuing the simplest possible choice, we will consider the case with $\psi_{\ell}$\,$=$\,$1$ and $\alpha_\ell = \alpha_0$; in this case, we can further set $\alpha_0=0$ without loss of generality. Intuitively, one can think of the nematic order parameter in Eq.~\eqref{OnsiteOrderParameter} as an anisotropic change of the overlap of the Wannier states of the microscopic graphene sheets. In fact, projecting a distortion of the graphene hopping matrix elements onto the Dirac cones of a single layer of graphene leads exactly to $(\rho_x,\eta_z\rho_y)$.

Comparing the nematically distorted Fermi surfaces in Fig.~\ref{fig:FS}(c) to those presented in panels (a) and (b), we find that the intravalley graphene nematic order also breaks the threefold rotational symmetry of the moir\'e superlattice. However, it breaks $C_3$ symmetry in a different way than the moir\'e nematic order shown in panel (b), namely, by shifting the Dirac cones of the individual graphene layers away from the $\bar{K}$ and $\bar{K}'$ points. As a result, the intravalley nematic order parameter, to a good approximation, rigidly shifts the Fermi surfaces in momentum space \textit{without} distorting them, in contrast to the case of moir\'e nematic order. Note that, to make the Fermi surface distortions in panels (b) and (c) comparable, we had to set the magnitude of the intravalley graphene nematic order parameter more than one order of magnitude larger than that of the moir\'e nematic order parameter. The reason for this discrepancy arises from the different form factors by which the nematic order parameter couples to the electronic degrees of freedom. Therefore, comparing $\Phi$ for different types of nematic order must be done with caution. Here, we choose to set the values of $\Phi$ such that similar amplitudes of Fermi surface distortions are obtained. Similar considerations apply for the case of intervalley graphene nematic order below.

\subsection{Intervalley graphene nematic order}
\label{sec:IVC}

Next, we look into the intervalley coherent (IVC) version of the graphene nematic order parameter. Since U(1)$_v$ and $C_3$ symmetries are simultaneously broken by intervalley nematic order, we have to describe it as a ($2\times 2$) matrix-valued order parameter $\Phi_{ij}$, with its first (second) index transforming as a vector under U(1)$_v \simeq $ O(2) (under $C_3$). This also has consequences for the allowed orientations of the nematic director in the absence of a displacement field, as we will explain in \secref{PhenomenologicalAnalysis}. 
Specifically, the symmetry representations in matrix notation are
\begin{equation}
\text{U(1)}_v:\quad \Phi \, \rightarrow \, R^T(\vartheta)\,\Phi, \quad C_3:\quad \Phi \, \rightarrow \, \Phi\, R\,(2\pi/3),
\end{equation}
where $R_{j,j'}(\vartheta)$\,$=$\,$ e^{\varepsilon_{j,j'}\vartheta}$ (with Levi-Civita symbol $\varepsilon_{j,j'}$) is the matrix representation of the rotation of a two-dimensional vector by angle $\vartheta$. 
Focusing, as above, on a local order parameter, \equref{GeneralFormOfNematicOrder} becomes
\begin{equation}
        \mathcal{H}^{\textsc{ivc}}_{\vec{\Phi}} = \hspace*{-0.1cm}\int_{\vec{r}}\, \sum_{i,j=1,2}\hspace*{-0.1cm}\Phi^\pdagger_{ij}(\phi^\pdagger_{ij})^{}_{_{\ell, s,\eta;\ell', s',\eta'}}(\vec{r}) \, c^\dagger_{\sigma,\ell, s, \eta}(\vec{r}) c^\pdagger_{\sigma,\ell', s', \eta'}(\vec{r}). \label{HPhiCoupl3}
\end{equation}
To further constrain the microscopic form of $\phi_{ij}$, we once again analyze $\vec{r}$-independent and layer-diagonal order parameters,
\begin{subequations}\begin{equation}
	(\phi^\pdagger_{ij})_{_{\ell, s,\eta;\ell', s',\eta'}}(\vec{r}) = 	\delta^\pdagger_{\ell,\ell'}\psi^\pdagger_{\ell}(\varphi^\pdagger_{ij})_{_{s,\eta;s',\eta'}}.
\end{equation}
In this case, there is only one remaining explicit form for a nematic order parameter: 
\begin{equation}
\label{eq:layerIVC}
\varphi^\pdagger_{ij} = \begin{pmatrix}\eta^{}_x & \eta^{}_y \rho_z \\ \eta^{}_y & -\eta^{}_x \rho^{}_z \end{pmatrix}_{ij}, \qquad \psi^{}_1=\psi^{}_4=0.
\end{equation}\label{IntervalleyCoherentOP}\end{subequations}
Interestingly, symmetry (together with the intralayer nature of the order parameter) requires the order parameter to be restricted to the two inner layers, $\ell$\,$=$\,$2,3$. Note that a layer-dependent valley rotation, i.e., $\eta_{x,y} \rightarrow e^{i\alpha_\ell \eta_z}\eta_{x,y}$ in \equref{IntervalleyCoherentOP}, is, in principle, allowed. However, both the angles $\alpha_\ell$ and the relative weight, $\psi_3/\psi_2$, are \textit{a priori} unknown; thus a natural starting point is to set $\psi_2=\psi_3=1$ and $\alpha_2=\alpha_3$ ($=0$ without loss of generality).

In order to study the impact of this type of nematicity, we first need to determine the possible stable configurations of $\Phi_{ij}$. Using the complex-vector notation, $\hat{\Phi}_j$\,$\equiv$\,$\Phi_{j1}$\,$+$\,$ i\,\Phi_{j2}$, the most general free energy is 
\begin{equation}
   \mathcal{F}^\pdagger_{\text{IVC}} =  a \hat{\vec{\Phi}}^\dagger \hat{\vec{\Phi}}+b^\pdagger_1 |\hat{\vec{\Phi}}^\dagger \hat{\vec{\Phi}}|^2 + b^\pdagger_2 |\hat{\vec{\Phi}}^\mathrm{T}\hat{\vec{\Phi}}|^2 + \mathcal{O}(\hat{\vec{\Phi}}^6) \label{FreeEnergyExpIVCNematic}
\end{equation}
up to fourth order in $\hat{\vec{\Phi}}$. Minimizing this free energy, we obtain two U(1)$_v$-inequivalent solutions, $\hat{\vec{\Phi}}$\,$=$\,$(1,\pm i)^\mathrm{T}$ and $\hat{\vec{\Phi}} = e^{2 i \varphi }(1,0)^\mathrm{T}$. In matrix form, they are then expressed as:
\begin{equation}
\Phi^\pdagger_{ij} = \begin{pmatrix} 1 & 0 \\ 0 & \pm 1 \end{pmatrix}_{ij}, \qquad \Phi^\pdagger_{ij} = \begin{pmatrix} \cos2 \varphi & \sin2 \varphi \\ 0 & 0 \end{pmatrix}_{ij}. \label{CouplingIVC}
\end{equation}
While the first minimum in Eq.~\eqref{CouplingIVC} preserves $C_3$ [modulo U(1)$_v$] and hence, cannot account for the experimental reports of threefold rotational symmetry breaking, the second one does indeed break it. The reason is that only the second option in Eq.~\eqref{CouplingIVC} can lead to any $C_3$-breaking signal in U(1)$_v$-symmetric variables such as the spectrum and the LDOS.

Plugging this solution into \equref{HPhiCoupl3} and reintroducing $\vec{\Phi} = (\cos 2 \varphi,\sin 2 \varphi)$, the term in the Hamiltonian corresponding to the presence of IVC graphene nematic order is given by
\begin{equation}
     \mathcal{H}_{\vec{\Phi}}^{\text{IVC}} = \int_{\vec{r}} \sum_{\ell=2,3}  \vec{\Phi}\cdot \begin{pmatrix}(\eta_x)_{\eta,\eta'} \\ (\eta_y)_{\eta,\eta'}(\rho_z)_{s,s}  \end{pmatrix}  c^\dagger_{\sigma,\ell, s, \eta}(\vec{r}) c^\pdagger_{\sigma,\ell, s, \eta'}(\vec{r}), \label{InterValleyGrapheneNematic}
\end{equation}
where summation over repeated indices is implied; the explicit sum over $\ell$ is used to indicate that only $\ell = 2,3$ contribute. 

In \figref{fig:FS}(d), we plot the distortions of the Fermi surfaces resulting from the intervalley graphene nematic order in \equref{InterValleyGrapheneNematic}. Note that, in this case, because valley U(1)$_v$ symmetry is broken, it is not possible to associate a Fermi surface with a single valley. As a result, even the unperturbed Fermi surfaces in Fig.~\ref{fig:FS}(d) (dashed lines) look different at first sight from those in panels (b), (c). When viewed as a two-valley model, the set of occupied electronic states (at zero temperature) in reciprocal space is formally given by the union 
$\mc{A} \equiv \mc{A}_+\cup \mc{A}_-$, where $\mc{A}_\pm $\,$\equiv$\,$ \{\vec{k} \in \mathrm{FBZ} \,\vert\, E_{n,\vec{k},\pm} - \mu \le 0\}$ defines the filled states in each valley; the boundary $\partial\mc{A}$ then delineates the Fermi surface. Thus, the form of the unperturbed Fermi surfaces shown in (d) can be regarded as a hybridization of the Fermi surfaces corresponding to different valleys shown in (a). Interestingly, the strongest effect of the intervalley graphene nematic order parameter in \equref{InterValleyGrapheneNematic} occurs in the pocket around the $\Gamma$ point in Fig.~\ref{fig:FS}(d). Furthermore, we observe that intervalley graphene nematic order does not translate to just a rigid shift of the Fermi surfaces, as opposed to its intravalley counterpart.

\section{Local density of states}
\label{sec:LDOS}

\begin{figure}[t]
\includegraphics[width=\linewidth]{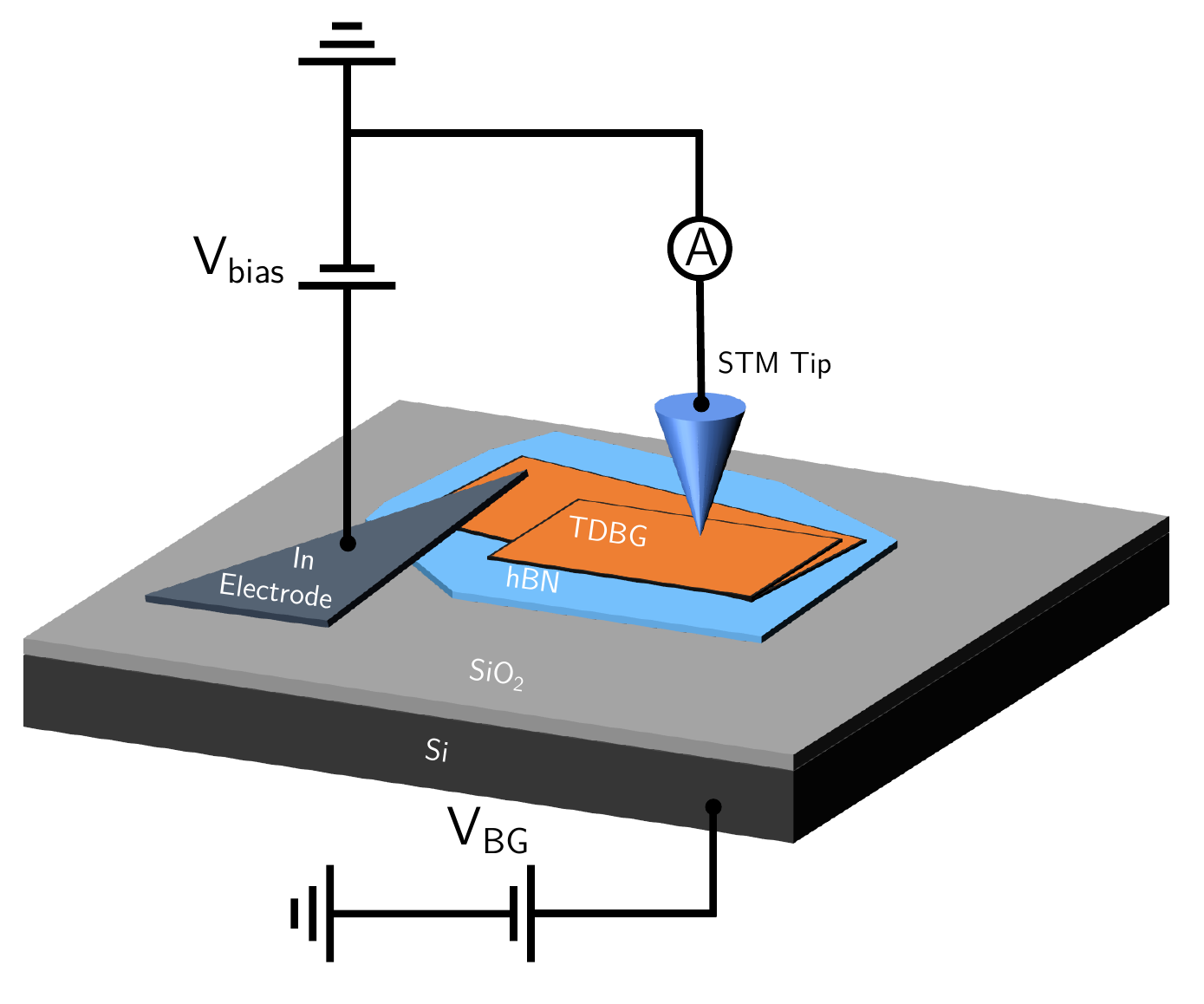}
\caption{\label{fig:STM_Device}Schematic device geometry for STM experiments. $V_\mathrm{bias}$ and $V_\textsc{bg}$ denote the bias voltage, and the back-gate voltage, respectively. The tDBG sample being probed is depicted in orange and consists of two bilayers of graphene---twisted relative to each other---on a hBN substrate. The structure is placed on top of a Si/SiO$_2$ chip and scanning tunneling microscopy and spectroscopy measurements are carried out using an atomically sharp tip.
}
\end{figure}

A promising avenue to probe the nature of the nematic order parameter in tDBG is provided by STM measurements. This technique, which has been widely employed in moir\'e systems \cite{PasupathySTM,YazdaniSTM,AndreiSTM,NadjPergeSTM,STMReview,zhang2020visualizing,liu2020spectroscopy}, consists of scanning a sharp conducting tip across the sample and recording the variation of the ensuing tunneling current $I$ as a function of the applied bias voltage $V$ for different positions of the tip (Fig.~\ref{fig:STM_Device}). One can then locally extract the quantity $\mathrm{d}I/\mathrm{d}V$, which is expected to be proportional to the system's local density of states (LDOS). Extensive and detailed real-space $\mathrm{d}I/\mathrm{d}V$ maps as a function of electronic filling and energy were reported in \refcite{rubioverdu2020universal} for tDBG, revealing threefold rotational-symmetry breaking near half-filling of the CFB and strongly peaked at energies corresponding to the VFB.

In order to elucidate whether the STM data can be used to constrain the form of the nematic order parameter, we theoretically calculate the LDOS in tDBG for each of the three types of nematic order discussed in the previous section. To compute the LDOS within the continuum model, we note that, in momentum space, the interlayer coupling hybridizes the eigenstates at a Bloch vector $\vec{k}$ in the moir\'{e} Brillouin zone with those at $\vec{q}$\,$=$\,$\vec{k}$\,$+$\,$\vec{G}$, where $\vec{G}$\,$=$\,$m_1  \vec{G}^{\textsc{m}}_1 + m_2  \vec{G}^{\textsc{m}}_2$ for $m_1, m_2 \in \mathbb{Z}$, $\vec{G}^{\textsc{m}}_{1,2}$ being the moir\'{e} reciprocal lattice vectors defined by $\vec{a}^{\textsc{m}}_j \cdot \vec{G}^{\textsc{m}}_{j'} = 2 \pi \delta_{j,j'}$. Therefore, the continuum-model wavefunctions have the form
\begin{equation}
    \Psi^{}_n (\vec{k}) = \left(\mc{U}^{}_{n,\vec{k}} (\vec{G}^{}_1), \mc{U}^{}_{n,\vec{k}} (\vec{G}^{}_2), \ldots, \mc{U}^{}_{n,\vec{k}} (\vec{G}^{}_i),\ldots\right)^\textsc{t}.
\end{equation}
While the set of reciprocal vectors $\vec{G}_{i}$ is formally infinite, in practice, it is truncated so as to retain a limited (but large) number of $\vec{q}$ points within a certain cutoff radius for the numerical diagonalization of the Hamiltonian (see Appendix~\ref{app:model}). Since the model is trivial in spin space (due to the negligible SOC) and diagonal in valley space (due to the small intervalley hopping terms present for small twist angles), each $\mc{U}$ term can be thought of as an eight-component column vector in layer and sublattice space:
\begin{equation} \label{eq_U}
  \mc{U}^{}_{n,\vec{k}} (\vec{G}) =   \left(U^{A_1}_{n,\vec{k}} (\vec{G}), U^{A_2}_{n,\vec{k}} (\vec{G}), \ldots, U^{B_4}_{n,\vec{k}}(\vec{G}) \right)^\textsc{t}.
\end{equation}
Focusing first on the cases where U(1)$_v$ symmetry is also preserved in the presence of nematic order, the LDOS per spin and valley is then given by \cite{peres2009local}
\begin{alignat*}{1}
\mc{D} \, (\boldsymbol{r}, E) &= \hspace*{-0.2cm} \sum_{\substack{n, \boldsymbol{k}\\\boldsymbol{G}, \boldsymbol{G}'}} e^{-i (\boldsymbol{G}-\boldsymbol{G}')\cdot \boldsymbol{r}}\, \mc{U}^\dagger_{n, \boldsymbol{k}} (\boldsymbol{G}')\, \mc{U}^{\pdagger}_{n, \boldsymbol{k}} (\boldsymbol{G})\, \delta (E- E^{}_{n, \boldsymbol{k}}).
\end{alignat*}
However, due to the experimental setup of the STM measurements, the electrons on the STM tip are expected to tunnel primarily to the topmost graphene layer, $\ell$\,$=$\,$4$. Accordingly, in order to make reasonable comparisons with the experimental data, we project the LDOS onto the topmost of the four graphene sheets. We obtain:
\begin{alignat}{1}
\label{eq:ldos}
&\mc{D}' (\boldsymbol{r}, E) = \sum_{n, \boldsymbol{k}} \sum_{\boldsymbol{G}, \boldsymbol{G}'} e^{-i (\boldsymbol{G}-\boldsymbol{G}')\cdot \boldsymbol{r}}\, \delta (E- E^{}_{n, \boldsymbol{k}})\\
\nonumber &\quad\times\left(\left[U^{A_4}_{n, \boldsymbol{k}} (\boldsymbol{G}')\right]^* U^{A_4}_{n, \boldsymbol{k}} (\boldsymbol{G}) 
+\left[U^{B_4}_{n, \boldsymbol{k}} (\boldsymbol{G}')\right]^* U^{B_4}_{n, \boldsymbol{k}} (\boldsymbol{G})\right). 
\end{alignat}
Equation~\eqref{eq:ldos} can be generalized in a straightforward way to the situation where U(1)$_v$ symmetry is broken---which is indeed the case for intervalley graphene nematic order---by extending the wavefunction $\mc{U}^{}_{n,\vec{k}}$ to a sixteen-component vector.

\begin{figure*}[htb]
\includegraphics[width=\linewidth]{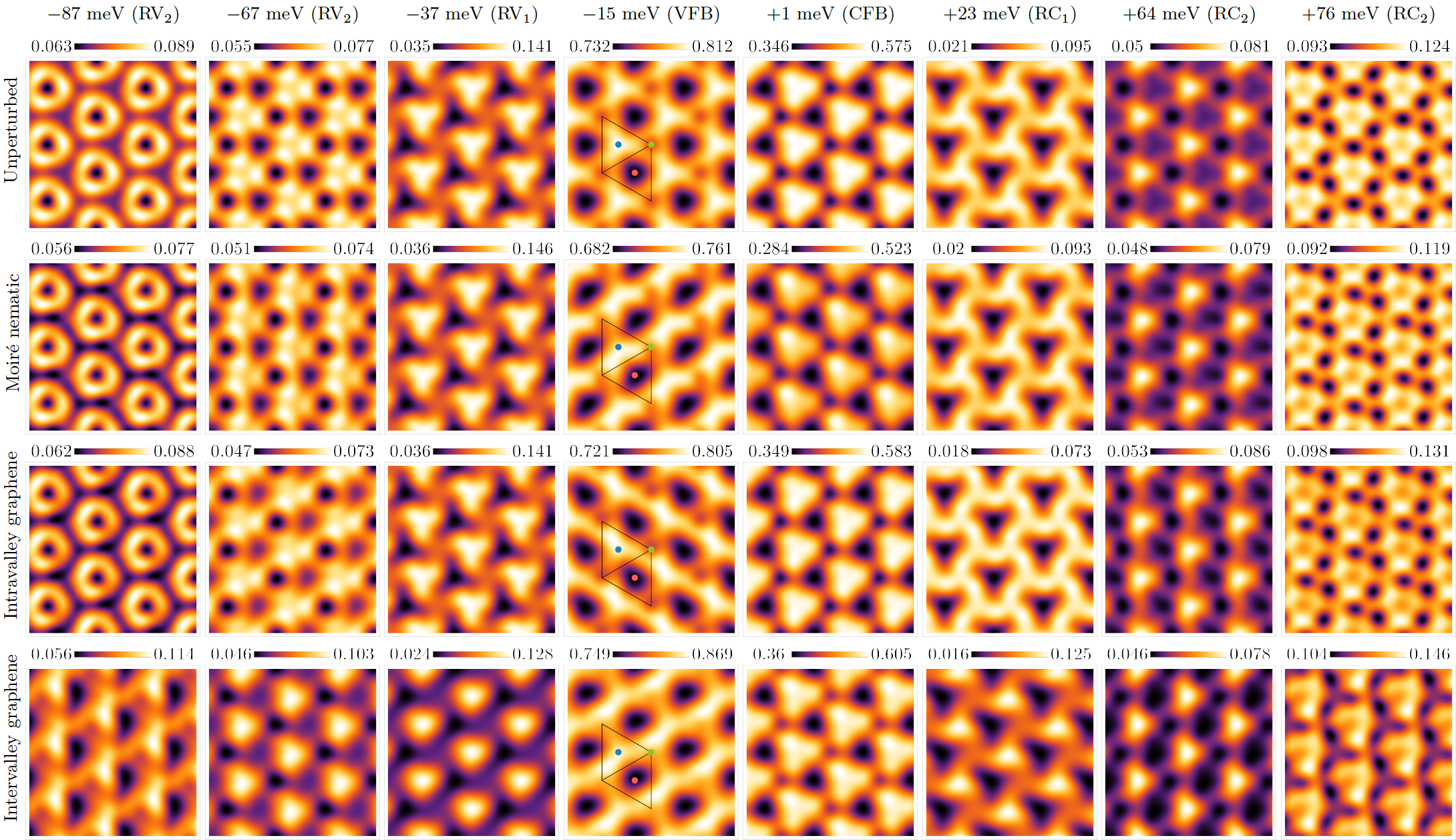} 
\caption{\label{fig:Maps}Real-space maps of the LDOS per spin (in arbitrary units) over a 20~nm\,$\times$\,20~nm region, computed at a filling fraction of $\nu$\,$=$\,$0.475$ without nematicity (labeled as unperturbed) as well as in the presence of moir\'{e} nematic, intravalley graphene nematic, and IVC (intervalley) graphene nematic orders. The magnitudes $\Phi$ used for the respective nematic order parameters are the same as in \figref{fig:FS}. The different energies at which these LDOS maps are computed are also indicated by the dashed horizontal lines in Fig.~\ref{fig:Lattice}. 
The black rhombus outlines the moir\'{e} unit cell, and the blue, green, and red dots mark the ABAB, BAAC, and ABCA sites, respectively.}
\end{figure*}

\begin{figure*}[htb]
\includegraphics[width=\linewidth]{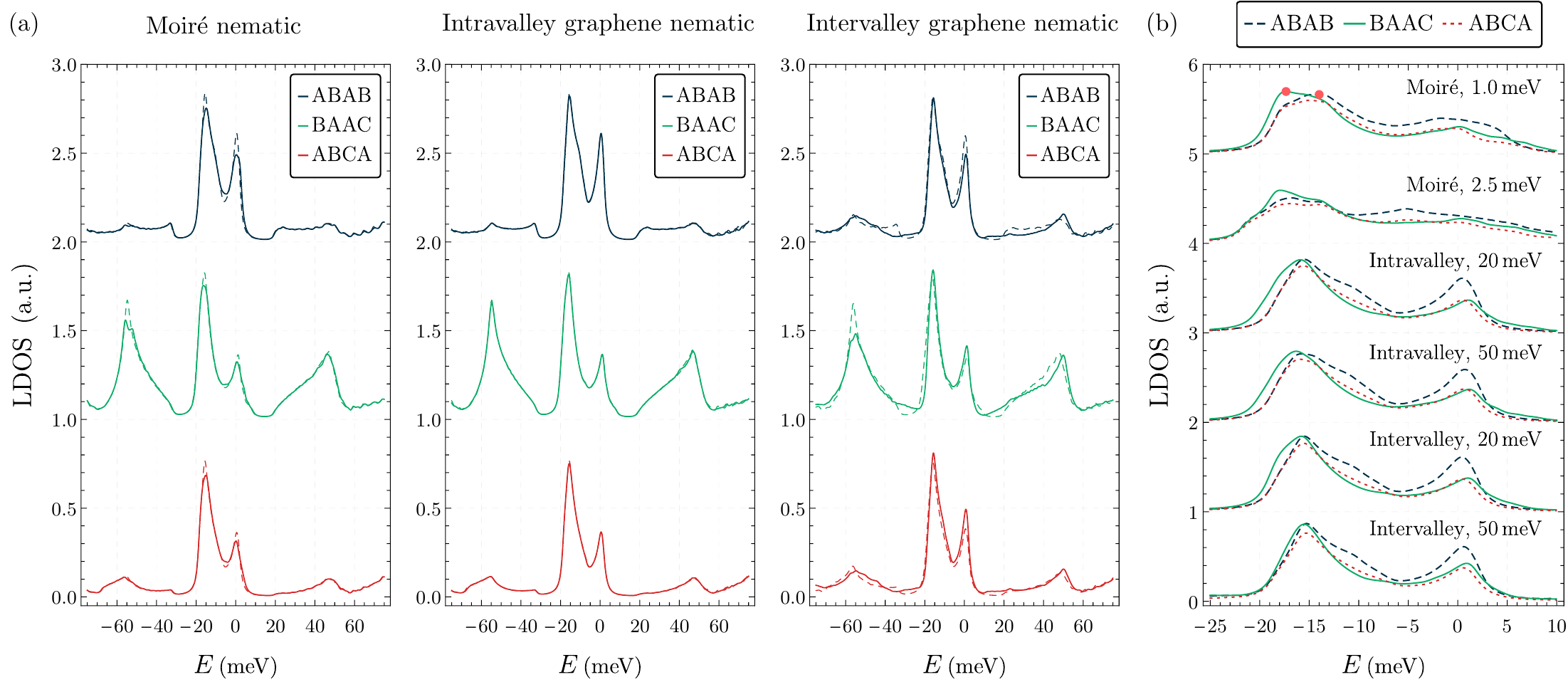}
\caption{\label{fig:LDOS} The LDOS (per spin) at the three inequivalent high-symmetry sites (insets in Fig.~\ref{fig:Maps}) as a function of energy ($E$) for the three different forms of nematic order. The curves, which are proportional to the theoretical prediction for the $\mathrm{d}I/\mathrm{d}V$ spectra in STM measurements, have been displaced vertically for the sake of visual clarity. The LDOS at each site in the absence of nematic order is represented by dashed lines. The band filling and the nematic order parameters' magnitudes were set to the same values as in Fig.~\ref{fig:FS}. (b) Zooming in on the spectrum near the Fermi level reveals that a site-dependent splitting of the VFB (valence flat band) peak, marked by the red dots, only occurs in the presence of moir\'{e} nematic order. Such a splitting is noticeably absent for the other two types of graphene nematic order regardless of the magnitude of the nematic order parameter. }
\end{figure*}

\subsection{Theoretical results and comparison with STM data}

We begin by computing the energy-dependent unperturbed LDOS maps (i.e., without nematicity), which are presented in the top row of Fig.~\ref{fig:Maps}. The band filling was set to $\nu$\,$=$\,$0.475$---corresponding to a chemical potential crossing the conduction flat band [CFB; see Fig.~\ref{fig:Lattice}(b)]---so as to have the same filling for which STM measurements detected threefold rotational-symmetry breaking \cite{rubioverdu2020universal}. Moreover, when including nematic order, the strengths of the order parameters were set to $\Phi$\,$=$\,$0.5$\,meV, $20$\,meV, and $50$\,meV for moir\'e nematic, intravalley graphene nematic, and intervalley graphene nematic orders, respectively. Our calculations and discussions in this section complement and extend the results previously presented in \refcite{rubioverdu2020universal}.

The maps for the valence flat band (VFB) and the conduction flat band (CFB), shown in the columns associated with the energies  $-15$~meV and $+1$~meV, respectively, reveal that the LDOS is mostly concentrated at the ABAB moir\'{e} sites for both bands (blue dots). This is in perfect agreement with the experimental observations of \refcite{rubioverdu2020universal} for filling fractions where nematic order is absent. On increasing the energy and moving away from the two flat bands, remarkably intricate triskelion structures emerge at the energies of the first set of remote bands around $-37$ and $+23$~meV (designated RV$_1$ and RC$_1$, respectively). These features, which were also observed experimentally, stand in contrast to the case of tBG, in which simpler circular bright spots corresponding to the AA stacking regions were reported \cite{PasupathySTM}. Moving away from these energies to the second set of remote bands (RV$_2$ and RC$_2$), the spatial distribution of the LDOS maps changes significantly, as the LDOS is now peaked at the BAAC sites---a feature also seen experimentally in \refcite{rubioverdu2020universal}.

In the presence of nematic order, the threefold rotational symmetry of the LDOS that is clearly visible in the top row of Fig.~\ref{fig:Maps} is patently broken. The most important feature is that the LDOS maps, for the energy corresponding to the VFB, now display prominent unidirectional stripes connecting the ABAB regions while the ABCA sites remain dark. 
The orientation of these stripes depends on the choice of the angle $\varphi$ of the nematic director and can be generically controlled by the displacement field, as we demonstrate in Sec.~\ref{sec:EField}. 
For concreteness, we have chosen $\varphi$\,$=$\,$0$ in Fig.~\ref{fig:Maps}, which yields stripes that develop nearly parallel to one of the moir\'{e} crystal axes for all three types of nematic order. 
Although the magnitude of the nematic order parameter cannot be directly compared among the three ans\"{a}tze, since their couplings to the unperturbed Hamiltonian are intrinsically different, the values of $\Phi$ were chosen so as to produce comparable changes in the Fermi surface across all three cases (refer to Fig.~\ref{fig:FS}). 

While the $C_3$-symmetry breaking manifests primarily at energies corresponding to the VFB, particularly for the cases of moir\'e and intervalley graphene nematic orders, the latter also strongly affects the LDOS at energies corresponding to the remote bands. In contrast, in the cases of intravalley graphene and moir\'e nematic orders, a much weaker twofold anisotropy is visible for the remote bands. Given that the STM $\mathrm{d}I/\mathrm{d}V$ maps reported in \refcite{rubioverdu2020universal} find that the signatures of threefold rotational-symmetry breaking are pronounced only at energies corresponding to the flat bands, being essentially absent at the remote bands, we conclude that intervalley graphene nematic order is unlikely to explain the nematicity experimentally observed in tDBG. Interestingly, we note that although the orientation of the nematic director is kept fixed in our calculations, the orientation of the twofold anisotropy seen in the LDOS maps varies depending on the energy being probed.

In addition to studying the LDOS as a function of position, 
we can obtain a complementary perspective into the symmetry-broken phase by looking at the energy-dependent LDOS at the high-symmetry sites (ABAB, BAAC, and ABCA). This is shown in Fig.~\ref{fig:LDOS}(a) for each of the three types of nematic order considered here, in comparison with the LDOS of the non-nematic state (dashed lines). Among the three ans\"{a}tze, the moir\'{e} nematic order parameter causes the strongest change of the LDOS at the energies corresponding to the VFB. In contrast, the changes caused by intravalley graphene nematicity are hardly discernible at the three high-symmetry sites, whereas for the intervalley nematic case, the strongest changes occur at the remote bands. We confirmed that these conclusions are not affected by varying the magnitudes $\Phi$ of the nematic order parameters. 

As mentioned previously, the STM data of \refcite{rubioverdu2020universal} found that the strongest changes caused by nematic order occur at the VFB. Thus, combined with the LDOS maps shown in Fig.~\ref{fig:Maps}, our results are consistent with and indicative of a moir\'{e} nematic order parameter, rather than either type of graphene nematic order. To further validate this conclusion, in Fig.~\ref{fig:LDOS}(b), we zoom in on the energy dependence of the LDOS at the three high-symmetry sites for energies corresponding to the valence flat band. It is clear that \textit{only} in the case of moir\'e nematic order, the peaks of the LDOS corresponding to the VFB move to different energies for different high-symmetry sites (red dots in the upper curve), in excellent agreement with the experimental observations \cite{rubioverdu2020universal}. For the other ans\"{a}tze, the LDOS peaks are coincident and located at essentially the same energy for the three sites' positions. The site-dependent LDOS peak splitting depicted in the upper curve correctly reproduces this important feature noted in the experimental STM $\mathrm{d}I/\mathrm{d}V$ spectrum. Therefore, we conclude that the most likely form of the nematic order parameter realized in tDBG is the moir\'e nematic.

\subsection{Importance of the form factors}

Many of our findings about the impact of the different forms of nematicity on the LDOS and on the Fermi surfaces are actually a consequence of the form factor encompassed by the matrix-valued function $\vec{\phi}$ in \equref{GeneralFormOfNematicOrder}. As such, several of these results can be understood qualitatively from the distribution of the spectral weight of the wavefunctions in layer and sublattice spaces (see Eq. (\ref{eq_U})). Let us define the spectral weights
\begin{equation}
\label{eq:weight}
P^{}_n (\ell, s) \equiv
\frac{1}{N^{}_{\boldsymbol{k}}} \sum_{\boldsymbol{k},\boldsymbol{G}}  \left[U^{s^{}_\ell}_{n, \boldsymbol{k}} (\boldsymbol{G})\right]^* U^{s^{}_\ell}_{n, \boldsymbol{k}} (\boldsymbol{G}), 
\end{equation}
where the summation over $\boldsymbol{k}$ runs over a grid of $N^{}_{\boldsymbol{k}}$ momentum points in the moir\'{e} Brillouin zone and $U^{s^{}_\ell}_{n, \boldsymbol{k}}$ as defined in Eq. (\ref{eq_U}). Intuitively, $P_n$ provides information about the momentum-averaged weight of the wavefunction corresponding to band $n$ in each layer $\ell$ and sublattice $s$. Note that normalization of the wavefunctions implies that $\sum_{\ell, s} P_n (\ell, s) = 1$ by definition. Therefore, $P_n$, for a fixed $n$, may also be regarded as a discrete probability distribution. 

\begin{figure}[tb]
    \centering
    \includegraphics[width=\linewidth]{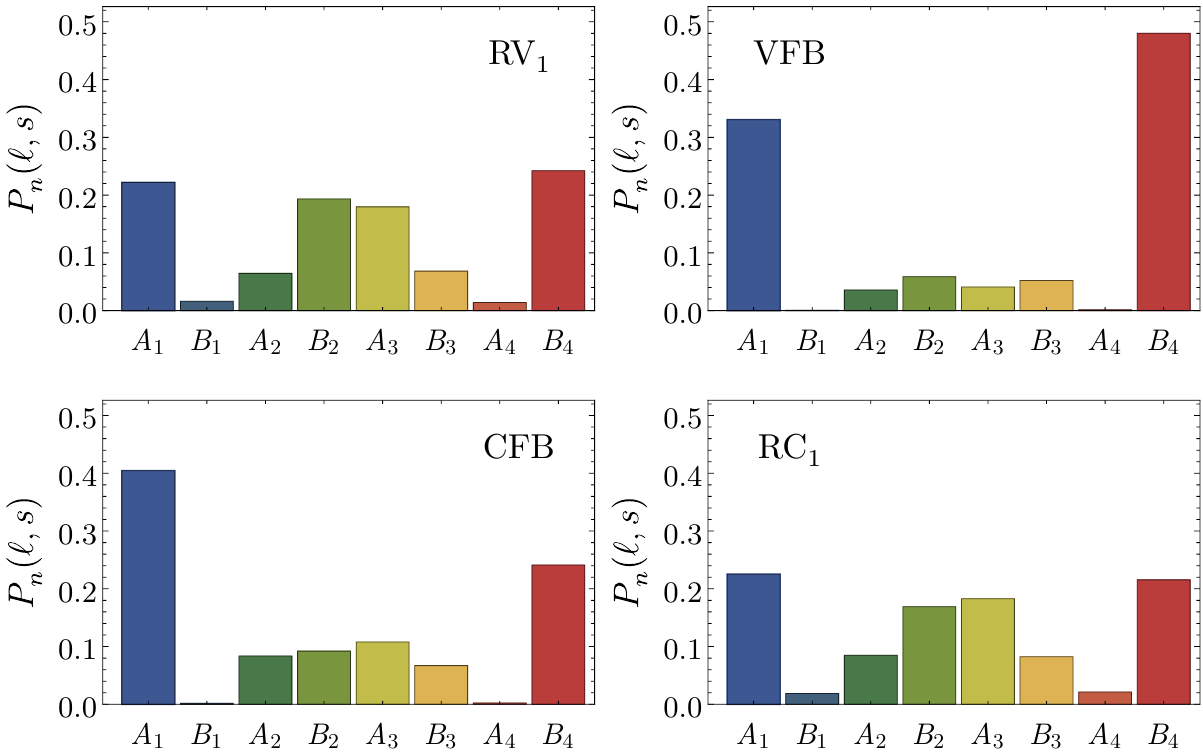}
    \caption{Momentum-integrated spectral weight $P^{}_n (\ell, s)$ of the wavefunction of each band $n$ in layer $\ell$ and sublattice $s$, as defined by Eq.~\eqref{eq:weight}. The VFB (CFB) has maximum spectral weight in the topmost (bottom-most) layer of the system, corresponding to the $B_4$\,($A_1$) sites. The histograms of the spectral weights of the remote bands, however, are more spread out across layer and sublattice space.}
    \label{fig:weight}
\end{figure}

Figure~\ref{fig:weight} displays the histograms of $P_n (\ell, s)$ associated with the two flat bands as well as the first valence and conduction remote bands. We see that the wavefunctions for the VFB (and to a lesser extent, the CFB) have a strong projection on the $B$ sublattice of the topmost ($\ell=4$) layer, whereas those for the remote bands are more spread out. As discussed above, the tunneling of the electrons from the STM tip to the tDBG device takes place via the topmost layer. Since the VFB is the band with the largest spectral weight on this layer, the projected LDOS as shown in \figref{fig:LDOS} and the $\mathrm{d}I/\mathrm{d}V$ spectra of STM measurements are naturally the largest for this band. It also makes the impact of the nematic order parameter most prominent in the VFB.

Furthermore, in the CFB and VFB, the wavefunctions are strongly polarized in one sublattice within each of the two dominant layers ($\ell$\,$=$\,$1,4$). This is consistent with our finding that the magnitude of the order parameter $\vec{\Phi}$ has to be much larger in the case of intravalley graphene nematic order ($\propto \rho_{x,y}$) than for the moir\'e nematic order ($\propto \rho_{0}$) so that we obtain comparable effects on the Fermi surface and the LDOS in the two cases.

The layer dependence of the spectral weight of the different bands also conveniently explains why intervalley coherent graphene nematicity yields more pronounced effects at the remote bands.
From  Eq.~\eqref{eq:layerIVC}, which describes the IVC nematic order parameter, we see that $\psi^{}_1$\,$=$\,$\psi^{}_4$\,$=$\,$0$, i.e., the nematic order parameter has no weight on the outer layers. As the spectral weight of the VFB (and also the CFB) is strongly peaked at the outer layers, the IVC nematic order parameter in Eq.~\eqref{IntervalleyCoherentOP} only weakly affects the LDOS at the VFB (and CFB). By the same token, a large magnitude $\Phi$ of the intervalley graphene nematic order parameter is required to induce a significant distortion of the Fermi surfaces in \figref{fig:FS}(d).

\section{Electric-field induced rotation of the nematic director}
\label{sec:EField}
In our analysis so far, the orientation $\varphi$ of the nematic director or, equivalently, the phase of the complex-valued nematic order parameter $\hat{\Phi}$\,$=$\,$  \Phi_1$\,$+$\,$i\, \Phi_2$\,$=$\,$|\hat{\Phi}|\,e^{2i\varphi}$, has been treated as a free parameter. In this section, we analyze the constraints on $\varphi$ imposed by the $C_3$ and $C_{2x}$ symmetries of the problem. In particular, we focus on how the breaking of $C_{2x}$ by the applied displacement field can be used to tune and control the orientation of the nematic director in tDBG. Before discussing the case of tDBG specifically, we start with a general phenomenological model suitable for any moir\'e superlattice with a lattice point group that contains $C_3$ and $C_{2x}$. 

\subsection{Phenomenological analysis}\label{PhenomenologicalAnalysis}
For our symmetry-based discussion, let us first focus on nematic order parameters that are valley-diagonal (such as the moir\'e nematic and the intravalley graphene nematic orders). In these cases, we can neglect the $\text{U(1)}_v$ symmetry in \equref{U1vSymmetry}, as the order parameter transforms trivially under it. Consequently, we only have to take into account the point-group symmetries, which form the group $D_3$ for tDBG  without a displacement field ($D$\,$=$\,$0$) [see Fig.~\ref{fig:Lattice}(a)].
As readily follows from the action of these symmetry operations on the nematic order parameter $\hat{\Phi} = \Phi_1 + i\, \Phi_2$ [see Eq.~\eqref{ActionOfSyms}], with components $\Phi_j$ transforming under the irrep $E$ of $D_3$, the free-energy expansion to fourth order is \cite{RafaelsPaperTBG,hecker2018vestigial,jin2019dynamical,little2020three}
\begin{equation}
    \mathcal{F} = \frac{a}{2} |\hat{\Phi}|^2 + \frac{\gamma}{3} \, \text{Re}(\hat{\Phi}^3) + \frac{u}{4} |\hat{\Phi}|^4 + \mathcal{O}(\hat{\Phi}^5). \label{FreeEnWithout}
\end{equation}
Note that the analogous third-order term $\text{Im}\,(\hat{\Phi}^3)$ transforms as the $A_2$ irrep of $D_3$ and is thus forbidden when $C_{2x}$ symmetry is present. Minimizing \equref{FreeEnWithout} yields two symmetry-inequivalent sets of director orientations, $\varphi_0=\pi/6,\pi/2,5\pi/6$, for $\gamma>0$, and $\varphi_0=0,\pi/3,2\pi/3$ for $\gamma<0$. These correspond to bond order along one of the three high-symmetry nearest-neighbor directions \cite{2019arXiv191111367F}. As a result, $\hat{\Phi}$ is a 3-state Potts-nematic order parameter. In both cases, $C_3$ and two of the three in-plane twofold rotations [dashed black lines in the inset of Fig.~\ref{fig:Lattice}(a)] are broken; as such, the residual point group is $C_2$ (generated by $C_{2x}$).

Once a displacement field is applied, $D \neq 0$, $C_{2x}$ is broken, and a term proportional to $\text{Im}(\hat{\Phi}^3)$ is allowed in \equref{FreeEnWithout}:
\begin{align}\begin{split}
    \mathcal{F}^{}_{D} &= \frac{a}{2} |\hat{\Phi}|^2 + \frac{\gamma}{3} \, \text{Re}(\hat{\Phi}^3) + \frac{\widetilde{\gamma}}{3} \, \text{Im}(\hat{\Phi}^3) + \frac{u}{4} |\hat{\Phi}|^4 + \mathcal{O}(\hat{\Phi}^5) \\
    &= \frac{a}{2} |\hat{\Phi}|^2 +  \frac{\gamma}{3} \, \text{Re}[(1-i\widetilde{\gamma}/\gamma)\hat{\Phi}^3] + \frac{u}{4} |\hat{\Phi}|^4 + \mathcal{O}(\hat{\Phi}^5), \label{FreeEnWithEx}
\end{split}\end{align}
where $\widetilde{\gamma}$ has to be an odd function of $D$, since $D$ transforms as the $A_2$ irrep of $D_3$. Of course, all other coefficients have to be even under $D$\,$\rightarrow$\,$-D$. As can be best seen from the second line of Eq.~\eqref{FreeEnWithEx}, the finite displacement field rotates the orientation of the nematic director away from the two distinct sets of high-symmetry orientations of $\varphi_0$ mentioned above. In particular, defining the director's rotation as $\Delta \varphi \equiv \varphi - \varphi_0$, we obtain
\begin{equation}
\label{eq:rotation}
    \Delta \varphi = \frac{1}{6} \, \text{arg}\left(\gamma + i  \, \widetilde{\gamma}\right). 
\end{equation}
While it readily follows from symmetry considerations that $\Delta \varphi\,(D)$ is an odd function of $D$, $\Delta \varphi\,(D)=-\Delta \varphi\,(-D)$, the strength of this effect can only be determined by a microscopic calculation. Clearly, the maximum possible rotation is $\Delta \varphi_{\mathrm{max}} = \pm \pi /12 = \pm 15^{\circ}$.

As anticipated above, this analysis has to be modified for intervalley nematic order, which breaks both $\text{U(1)}_v$ and the point-group symmetry at the same time and hence, is described by the matrix-valued order parameter $\Phi_{ij}$ introduced in \secref{sec:IVC}. Using the complex-vector representation, $\hat{\Phi}_j$\,$\equiv$\,$\Phi_{j1}$\,$+$\,$ i\,\Phi_{j2}$, the most general free energy up to fourth order is given by \equref{FreeEnergyExpIVCNematic}, which does not involve a third-order term, irrespective of whether $C_{2x}$ is present or not. Thus, the intervalley nematic order parameter is not a 3-state Potts order parameter. Due to the combination with $\text{U(1)}_v$ symmetry, the lowest-order terms that are sensitive to $\varphi$ in $\hat{\vec{\Phi}}$\,$=$\,$e^{2 i \varphi }(1,0)^\mathrm{T}$ are given by $g\, \text{Re} [(\hat{\vec{\Phi}}^\mathrm{T}\hat{\vec{\Phi}})^3]$ and $\widetilde{g}\, \text{Im} [(\hat{\vec{\Phi}}^\mathrm{T}\hat{\vec{\Phi}})^3]$. Since $C_{2x}$ implies $\widetilde{g}$\,$=$\,$0$, when $D=0$ we obtain two sets of six possible minima, $\varphi_0$\,$=$\,$n (\pi/6)$ and $\varphi_0$\,$=$\,$(\pi/12) + n (\pi/6)$, with $n=0,1,2,3,4,5$. Note that within each set, pairs of configurations are related by $\text{U(1)}_v$ and the order parameter manifold is isomorphic to $O(2) \times \mathbb{Z}_3$. As such, no true long-range intervalley nematic order is possible at finite temperature, although $\text{U(1)}_v$-symmetric combinations can order, such as the $C_3$-symmetry-breaking order parameter $\hat{\vec{\Phi}}^\mathrm{T}\hat{\vec{\Phi}}$. 

Once a finite displacement field is applied, $C_{2x}$ is broken and $\widetilde{g}$ becomes nonzero, leading to a continuous rotation of $\varphi$, analogous to \equref{eq:rotation}.  Of course, since the $\text{U(1)}_v$ symmetry is only an approximate symmetry, any small perturbation that breaks it will reduce the intervalley nematic order parameter to a 3-state Potts order parameter, as in the intravalley graphene and moir\'e nematic cases.  

\subsection{Microscopic analysis}
The phenomenological analysis presented above is very general and valid for any type of nematic order in an arbitrary triangular lattice. As such, it applies equally to tBG and to tDBG. What makes the latter system a more suitable candidate to observe the electric-field induced rotation of the nematic director is the fact that its band structure is strongly altered by the displacement field \cite{2019arXiv190108420R,2019arXiv190300852C,2019PhRvB..99g5127Z, FirstModel,2019arXiv190600623H}. Indeed, it is the displacement field that is responsible for ensuring a nonzero gap between the CFB and the VFB \cite{koshino2019band}. This suggests that the cubic coefficient $\widetilde{\gamma}$ (or $\widetilde{g}$) may be sizable and comparable to $\gamma$ (or $g$) in tDBG.

\begin{figure}[tb]
    \centering
    \includegraphics[width=\linewidth]{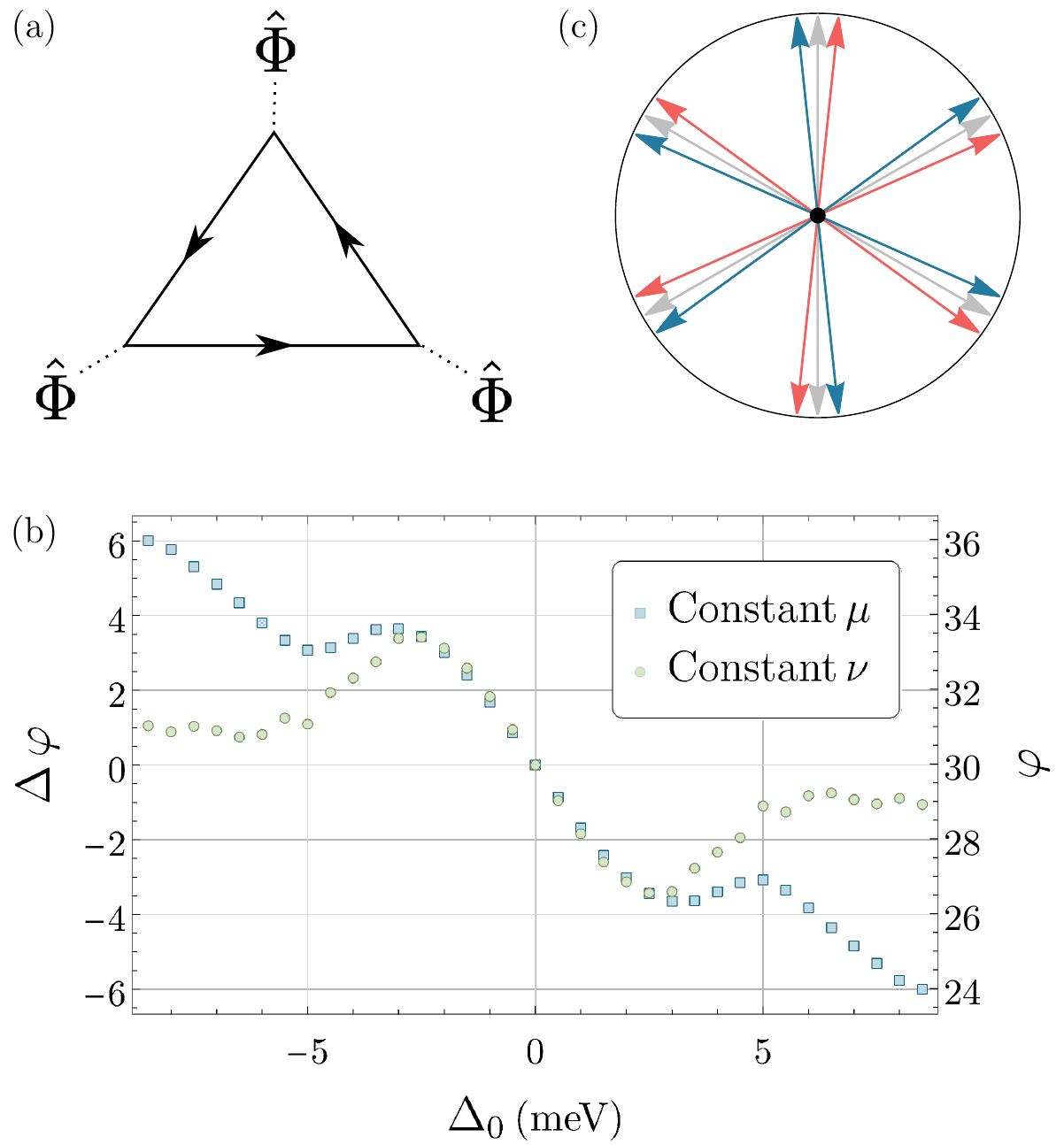}
    \caption{(a) One-loop Feynman diagram for the third-order contribution to the free energy in Eq.~\eqref{eq:thirdorder}. The solid black lines represent the electronic Green's function of the full continuum Hamiltonian with arbitrary $D$. (b) Variation of the angle of the nematic director, $\varphi$ (in degrees), and its rotation with respect to the high-symmetry direction, $\Delta\varphi = \varphi - 30^{\circ}$, as a function of the displacement field, calculated using Eq.~\eqref{eq:rotation} at $T$\,$=$\,$5$~K. The blue squares indicate the results for $\Delta\varphi$ calculated with the chemical potential $\mu$ held constant at the value shown in Fig.~\ref{fig:Lattice}(b), whereas the green circles denote the data obtained on working at a fixed filling of $\nu$\,$=$\,$0.475$. The parameter $\Delta_0$ measures the difference in electrostatic energy between adjacent graphene layers, such that the net perpendicular electric field across the sample is given by $D$\,$=$\,$\Delta_0/L$, where $L$ is the interlayer separation. (c) Orientations of the nematic director for $\Delta_0$\,$=$\,$-8.5$~meV (blue) and $+8.5$~meV (red), as specified by the endpoints of the constant-$\mu$ curve in panel (b). The symmetry-allowed values of $\varphi$ in the absence of a field are indicated in gray. }
    \label{fig:Rotation}
\end{figure}

\begin{figure*}
    \centering
    \includegraphics[width=\linewidth]{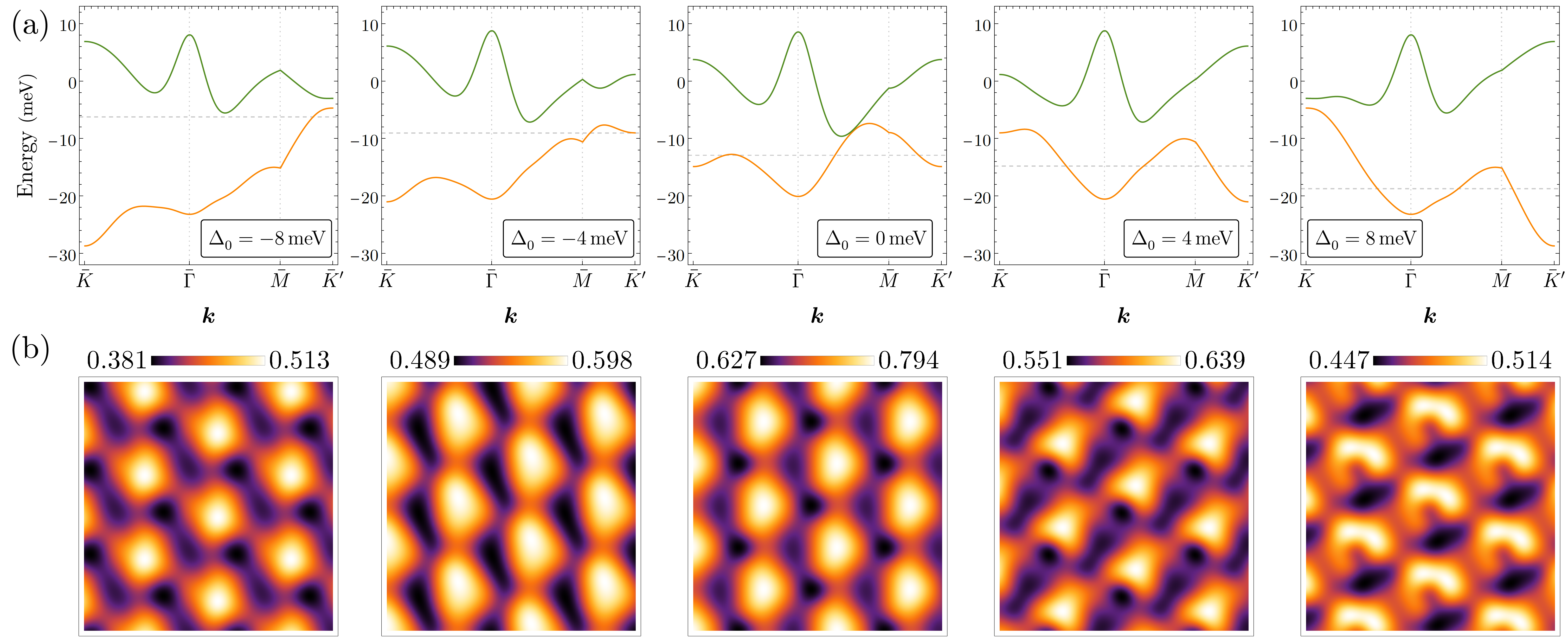} 
    \caption{(a) Band structure of tDBG for the $\eta$\,$=$\,$+$ valley at five different values of the displacement field $D=\Delta_0/L$. The band dispersions change appreciably with the field; note that without an applied electric field (central panel), the flat valence and conduction bands touch. The chemical potential is chosen  appropriately for each $\Delta_0$ so as to maintain a constant filling of $\nu$\,$=$\,$0.475$ of the CFB for all values of $D$. (b) Corresponding LDOS maps (in arbitrary units) after projecting onto the top layer, plotted over a 20~nm\,$\times$\,20~nm area, assuming a moir\'{e} nematic order parameter of strength $\Phi$\,$=$\,$0.5$~meV, as in Fig.~\ref{fig:Maps}. These LDOS maps are computed at the energies for which the stripes are visually most pronounced [marked as gray dashed lines in (a)].}
    \label{fig:stripes}
\end{figure*}

To confirm this hypothesis, we compute the $D$-induced rotation of $\varphi$ microscopically within the continuum model, concentrating on the case of moir\'e nematic order, which explains well the experimental STM data. Owing to the straightforward shift of the dispersion in Eq.~\eqref{ShiftOfTheBandstructureByOP} caused by this type of nematic order, the relevant coefficients $\gamma$ and $\widetilde{\gamma}$ in Eq.~\eqref{FreeEnWithEx} have particularly simple forms. We start from the microscopic expression of the free energy in the presence of both moir\'e nematic order and an electric field, 
\begin{equation}
    \Delta\mathcal{F} = -T \sum_{n,\vec{k}} \log \left(1 + e^{-(E^{}_{n,\vec{k}}-\mu + \vec{\Phi}\cdot\vec{f}(\vec{k}))/T} \right), \label{FreeEnergyForm}
\end{equation}
where $T$ denotes temperature, and we have neglected a term quadratic in $\vec{\Phi}$ that depends on the interaction but is irrelevant for our analysis. Here, $\vec{f}(\vec{k})$ is given by Eq.~\eqref{LowestOrderBasisFuncs}. Note that the sum over $n$ involves all bands, including spin and valley degrees of freedom, which just yields an overall factor of four in this expression and below. Expanding \equref{FreeEnergyForm} in $\vec{\Phi}$, one obtains the third-order contribution
\begin{equation}
\label{eq:thirdorder}
        \Delta \mc{F}^{}_3 = \hspace*{-0.25cm}\sum_{\substack{\mu_j=1,2\\n,\vec{k}}}\left[\prod_{j=1}^3\Phi^{}_{\mu_j}\right]  \frac{\sinh^4 \left(\frac{\xi^{}_{n,\vec{k}}}{2T}\right)}{3T^2\sinh^3 \left(\frac{\xi^{}_{n,\vec{k}}}{T}\right)} \left[\prod_{j=1}^3 f^{}_{\mu_j}(\vec{k})\right]
\end{equation}
with $\xi_{n,\vec{k}}$\,$\equiv$\,$E_{n,\vec{k}}-\mu$. Alternatively, one can arrive at this expression via a straightforward evaluation of the one-loop diagram in Fig.~\ref{fig:Rotation}(a). 
Comparing with Eq.~\eqref{FreeEnWithEx}, we find the coefficients 
\begin{align}
    \gamma + i  \, \widetilde{\gamma} = \frac{1}{4\,T^2} \sum_{n,\vec{k}} \frac{\sinh^4 \left(\frac{\xi^{}_{n,\vec{k}}}{2T}\right)}{\sinh^3 \left(\frac{\xi^{}_{n,\vec{k}}}{T}\right)}  \left[f^{}_1(\vec{k})+i f^{}_2(\vec{k})\right]^3. \label{ExpressionForRotation}
\end{align}
In agreement with our phenomenological analysis, we see that $\widetilde{\gamma}$\,$=$\,$0$ if $E_{n,\vec{k}}=E_{n,C_{2x}\vec{k}}$, which is precisely the case for $D=0$.

For the microscopic calculation of $\varphi$, we employ a continuum model in which the perpendicular electric field is constant and unscreened (unlike in Sec.~\ref{AFewNaturalMicroscopicForms}).
In Fig.~\ref{fig:Rotation}(b), we compute the rotation of the director $\Delta \varphi$, given by Eq.~\eqref{eq:rotation}, as a function of the (uniform) interlayer asymmetric potential $\Delta_0$. The latter is proportional to the displacement field as $D$\,$=$\,$\Delta_0/L$ (where $L$ is the vertical spacing between the layers), and enters the continuum model Hamiltonian as a difference in the electrostatic energy between adjacent layers (see Appendix~\ref{app:model} for details).  Since one can design an experimental setup either at a constant gate voltage, which fixes the chemical potential $\mu$, or at a constant filling fraction $\nu$ (of the CFB), which can be tuned by the voltages on the top and bottom graphite gates encasing the sample \cite{2019arXiv190306952S,ExperimentKim,PabllosExperiment}, we show results for both fixed $\mu$ (blue squares) and fixed $\nu$ (green circles). In either case, as expected, $\Delta \varphi$ is an odd function of $\Delta_0$, with pronounced nonlinear effects arising around $\Delta_0 \approx 3$ meV. Importantly, the rotation $\Delta \varphi$ for realistic values of $\Delta_0$ can be as large as about a third of the maximum possible rotation $\Delta \varphi_{\mathrm{max}} = 15^{\circ}$ [see also Fig.~\ref{fig:Rotation}(c)]. This demonstrates that the displacement field in tDBG provides a promising framework to control the electronic nematic director.

\subsection{Signatures in STM data}
Having established that the nematic director's rotation $\Delta \varphi$ is sizable for realistically accessible values of $\Delta_0$, we now investigate the experimental manifestations of this effect. An obvious candidate in this regard is the orientation of the stripes of the LDOS maps discussed in Fig.~\ref{fig:Maps} for the case of a nearly half-filled CFB with the tunneling energy set at the VFB. 

It is important to note, however, that there are actually \textit{two} closely related effects pertaining to the anisotropy of the LDOS when an electric field is applied. One effect is the rotation of the director discussed above. However, even if one were able to somehow keep the director fixed, as we theoretically did in Sec.~\ref{sec:LDOS} above, the anisotropy of the stripes in the LDOS could still change as a function of the electric field because the eigenfunctions and energies of the continuum Hamiltonian themselves are altered by the field. For this reason, we dub this change the ``band-structure effect'', to distinguish it from the effect of the rotation of the director. Note that a related band-structure effect was already present in Fig.~\ref{fig:Maps}, since the anisotropy of the stripes changed as a function of energy and of the type of nematic order considered. Thus, it is the combination of these two effects---the director's rotation and the band-structure effect---that will cause a net rotation of the stripes discerned in the LDOS maps for a changing displacement field.

In Fig.~\ref{fig:stripes}, we compute the LDOS maps as a function of $\Delta_0$, keeping the filling factor $\nu = 0.475$ constant---corresponding, approximately, to a half-filled CFB---while probing energies in the VFB. Figure~\ref{fig:stripes}(a) shows the significant changes in the band structure caused by the varying electric field; this is the origin of the band-structure effect discussed above. Simultaneously, we also self-consistently compute  the orientation of the director for each value of $\Delta_0$, as shown previously in Fig.~\ref{fig:Rotation}(b). The LDOS maps plotted in Fig.~\ref{fig:stripes}(b) clearly demonstrate that the stripes rotate considerably as a function of $\Delta_0$. Note that, to highlight this effect, we focused on the energies in the VFB that yield the sharpest signatures of the stripes [dashed lines in Fig.~\ref{fig:stripes}(a)]. As the field is varied from negative to positive values, the stripes rotate by an angle much larger than $\Delta \varphi$. This implies that the band-structure effect not only amplifies the field-induced rotation of the nematic director, but it is also the dominant contributor to the change in the LDOS anisotropy.

\subsection{Signatures in transport measurements}
\begin{figure}[tb]
\includegraphics[width=\linewidth]{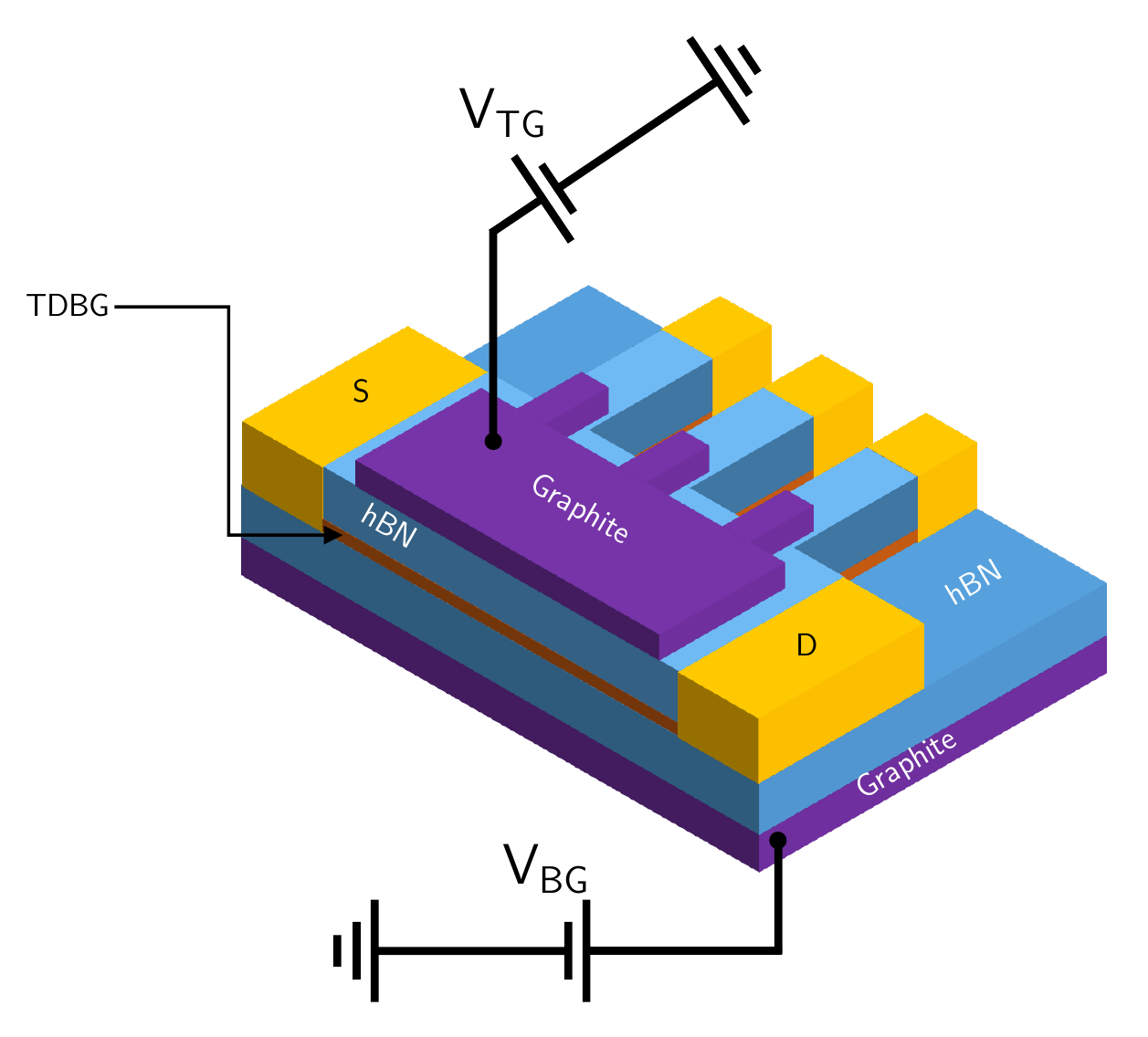}
\caption{\label{fig:Transport_Device} Illustrative sketch of a tDBG device for transport measurements. The setup of the gates allows one to control both the displacement field and the electronic density independently by varying the voltages on the top gate ($V_\textsc{tg}$) and the back gate ($V_\textsc{bg}$). {\sffamily S} and  {\sffamily D} represent the source and drain contacts, respectively. }
\end{figure}

A hallmark of nematic phases is the existence of a sizable resistivity anisotropy. While this phenomenon has been seen in experiments on tBG \cite{Cao2020_nematics}, experimental investigation of the resistivity anisotropy in tDBG remain to be performed, to the best of our knowledge. As opposed to STM setups, transport measurements \cite{PhysRevB.101.125428,PhysRevLett.125.176801,EnsslinExpDW} can usually be performed in the presence of top and back gates (Fig.~\ref{fig:Transport_Device}), which allow one to control both the displacement field and the electronic density independently \cite{2019arXiv190306952S,ExperimentKim,PabllosExperiment,burg2019correlated}. This provides more flexibility in stabilizing and tuning nematic order.

It is straightforward to derive, using symmetry considerations, the conductivity tensor $\sigma_{\alpha \beta}$ relating the in-plane electric current $J_\alpha$ and the electric field $\mc{E}_\beta$ ($\alpha,\beta$\,$=$\,$x,y$) via $J_\alpha$\,$=$\,$ \sum_\beta \sigma_{\alpha\beta} \mc{E}_\beta$. We find
\begin{equation}
    \sigma = \begin{pmatrix} \sigma^{}_0 + \lambda & \delta \\ \delta & \sigma^{}_0-\lambda \end{pmatrix}, \label{FormOfCondTensor}
\end{equation}
where $\lambda$ can only be nonzero if $C_3$ is broken, i.e., in the presence of nematic order, whereas $\delta\neq0$ requires both $C_3$ and $C_{2x}$ to be broken. It follows that $\sigma^{\mathrm{T}}$\,$=$\,$\sigma$ due to the assumption that time-reversal symmetry is preserved (in accordance with the Onsager relations). We emphasize that nematic order is essential to obtain a nontrivial conductivity tensor, $\sigma \neq \mathds{1}\sigma_0$.

To experimentally extract the two nontrivial components of the conductivity in the nematic phase, we propose a strategy employed in previous experimental works \cite{NematicOrderCuprates}. In particular, we consider a current applied along a direction that makes an angle $\Omega$ with respect to a high-symmetry crystalline axis, taken here to be the axis of the $C_{2x}$ rotational symmetry. In principle, $\Omega$ can be varied in discrete intervals by using a so-called ``sunbeam" pattern of Hall bars, as explained in \refcite{NematicOrderCuprates}. Writing $\vec{\mathcal{E}} = \sum_{j=1,2}\mathcal{E}_j \vec{e}_j(\Omega)$, and similarly for $\vec{J}$, where $e_1(\Omega) = (\cos \Omega,\sin \Omega)^{\mathrm{T}}$ and $e_2(\Omega) = (-\sin \Omega,\cos \Omega)^{\mathrm{T}}$, we can define the longitudinal resistivity, $\rho_L(\Omega)$\,$\equiv$\,$\mc{E}_1/J_1$, and the transverse resistivity, $\rho_T(\Omega)$\,$\equiv$\,$\mc{E}_2/J_1$.   
From \equref{FormOfCondTensor}, we then have
\begin{subequations}
\begin{align}
    \rho^\pdagger_L(\Omega) & = \frac{1}{\sigma^{}_0} - \frac{1}{\sigma_0^2} \left( \lambda \cos 2\Omega +\delta \sin 2\Omega\right), \\
    \rho^\pdagger_T(\Omega) & = \frac{1}{\sigma_0^2} \left( \lambda \sin 2\Omega -\delta \cos 2\Omega\right). \label{TransverseResistivity}
\end{align}\end{subequations}
We see that both the longitudinal and transverse resistivities show oscillatory behaviors as a function of $\Omega$ if and only if $C_3$ is broken. Due to the additional $\Omega$-independent (and likely dominant) contribution in $\rho_L$, we expect $\rho_T$ to be better suited for the experimental identification and study of nematic order \cite{NematicOrderCuprates}.

While it is clear by symmetry that $\delta$, and hence, the ``phase offset'' of the twofold oscillations of $\rho_T(\Omega)$, will depend on the displacement field as well as the orientation of the nematic director, a microscopic calculation is required to establish a direct relation. To this end, we employ a Boltzmann transport approach in the relaxation-time approximation \cite{BoltzmanTransportTheo}. Focusing, for notational simplicity, on the bands that lead to Fermi surfaces [one band for each valley $\eta=\pm$ and for each spin, denoted by CFB in Fig.~\ref{fig:Lattice}(a)], we obtain
\begin{subequations}\begin{equation}
    \sigma^\pdagger_{\alpha\beta} = \frac{2e^2 \tau}{\mc{V}} \sum_{\eta=\pm} \left\langle(\vec{v}^\pdagger_{\vec{k},\eta})^\pdagger_\alpha (\vec{v}^\pdagger_{\vec{k},\eta})^\pdagger_\beta\right \rangle_\eta, \label{GeneralExpressionForSigma}
\end{equation}
where all the nonuniversal properties are encoded in the generalized average
\begin{equation}
    \braket{\dots}_\eta \equiv \frac{4}{T N}\sum_{\vec{k}} \frac{\tau^\pdagger_{\vec{k},\eta}/\tau}{\cosh^2\hspace{-0.1em} \left(\frac{E^{\Phi}_{\vec{k},\eta}-\mu}{2T}\right)} \dots \label{DefinitionOfExpectationValue}
\end{equation}\label{FullExpressionForSigma}\end{subequations}
and in the velocities $\vec{v}_{\vec{k},\eta}$\,$=$\,$ \vec{\nabla}_{\vec{k}} E^{\Phi}_{\vec{k},\eta}$. Here, $N$ is the number of moir\'e unit cells, $\mc{V}$ is the area of the moir\'e unit cell, $E^{\Phi}_{\vec{k},\eta}$ is the band dispersion in the presence of nematic order, and $\tau_{\vec{k},\eta}$ is the relaxation-time scattering rate for momentum $\vec{k}$ in valley $\eta$. We further defined the average scattering rate $\tau$\,$=$\,$N^{-1} \sum_{\vec{k}} \tau_{\vec{k},\eta}$, which is the same in both valleys, as a consequence of time-reversal symmetry.

For moir\'e nematicity, the impact of nematic order is readily computed since $E^{\Phi}_{\vec{k},\eta}$\,$=$\,$E_{\vec{k},\eta}$\,$+$\,$\vec{\Phi}\cdot\vec{f}(\vec{k})$ [see Eq.~\eqref{ShiftOfTheBandstructureByOP}], and thus, $\vec{v}_{\vec{k},\eta}$\,$\rightarrow$\,$ \vec{v}^0_{\vec{k},\eta}$\,$+$\,$\sum_j \Phi_j (\vec{\nabla} f_j)(\vec{k})$, where $\vec{v}^0_{\vec{k},\eta}$ is the velocity without nematicity. Expanding \equref{FullExpressionForSigma} to linear order in $\vec{\Phi}$, we obtain
\begin{equation}
    \sigma^\pdagger_{\alpha\beta} = \delta^\pdagger_{\alpha,\beta}\, \sigma^\pdagger_0 + \frac{2e^2 \tau}{\mc{V}} \sum_j \mc{T}^{(j)}_{\alpha\beta}\, \Phi^\pdagger_j,
\end{equation}
where $\sigma_0 = (2e^2 \tau)\sum_\eta \braket{(\vec{v}^0_{\vec{k},\eta})_x (\vec{v}^0_{\vec{k},\eta})_x}_{\eta,0}/\mc{V}$ is the conductivity without nematic order, and we have introduced the third-rank tensor
\begin{alignat}{1}
    \mc{T}^{(j)}_{\alpha\beta} &= \sum_{\tau=\pm}\left[\left \langle \left(\vec{v}^0_{\vec{k},\eta}\right)^\pdagger_\alpha \partial^\pdagger_{k_\beta}f^\pdagger_j(\vec{k})\right\rangle^\pdagger_{\eta,0} + (\alpha \leftrightarrow \beta)\right] \\ 
    \nonumber & \quad -\frac{1}{T}  \left\langle\left(\vec{v}^0_{\vec{k},\eta}\right)^\pdagger_\alpha\left(\vec{v}^0_{\vec{k},\eta}\right)^\pdagger_\beta f^\pdagger_j(\vec{k}) \tanh\frac{E_{\vec{k},\eta}-\mu}{2T}\right\rangle_{\hspace{-0.2em}\eta,0}.
\end{alignat}
The extra subscript $0$ in the average indicates that it is defined as in \equref{DefinitionOfExpectationValue} but with respect to the bare dispersion $E_{\vec{k},\eta}$. We see that $\mc{T}$ is symmetric in two of the three indices, $\mc{T}^{(j)}_{\alpha\beta}=\mc{T}^{(j)}_{\beta\alpha}$, in consistency with the Onsager relations. We make the very natural and common assumption that the impurities preserve the point group symmetries \textit{on average}. Then, $\mc{T}$ transforms as a vector under $C_3$ and $C_{2x}$ in all of its indices, which reduces the number of independent components to two. Choosing them to be $\mc{T}_{xx}^{(1)}$ and $\mc{T}_{xx}^{(2)}$, we recover the form of $\sigma$ in \equref{FormOfCondTensor} with
\begin{subequations}\begin{align}
    \lambda &= \frac{2e^2 \tau}{\mc{V}} \left( \mc{T}^{(1)}_{xx} \Phi^\pdagger_1 + \mc{T}^{(2)}_{xx} \Phi^\pdagger_2 \right), \\
    \delta &= \frac{2e^2 \tau}{\mc{V}} \left( \mc{T}^{(2)}_{xx} \Phi^\pdagger_1 - \mc{T}^{(1)}_{xx} \Phi^\pdagger_2 \right).
\end{align}\end{subequations}
Defining $\mc{T}^{(1)}_{xx} + i \, \mc{T}^{(2)}_{xx} \equiv |t| e^{i\zeta}$, the transverse resistivity in \equref{TransverseResistivity} acquires the simple form
\begin{equation}
    \rho^\pdagger_T(\Omega) = \frac{2e^2 \tau}{\mc{V}\,\sigma_0^2}\, |t|\, \Phi^\pdagger_0 \sin(2\Omega + 2\varphi - \zeta), \label{FinalFormOfRhoT}
\end{equation}
which is the central result of this section. In accordance with our symmetry-based discussion above, we see that $\rho_T$\,$\propto$\,$\Phi_0$, i.e., it is only nonzero in the presence of nematic order. Hence, the transverse resistivity provides an excellent tool to probe the latter, even if only one or a few generic $\Omega$ are accessible experimentally \cite{Cao2020_nematics}. Interestingly, the phase offset of the twofold oscillations of $\rho_T(\Omega)$ also contains information about the degree of $C_{2x}$ breaking. As with the LDOS, there are two complementary effects---the orientation ($\varphi$) of the nematic director and the displacement-field-induced deformation of the band structure (related to the aforementioned band-structure effect) that breaks $C_{2x}$ symmetry, parametrized here by $\zeta$. Note that $\zeta=0$ for $D=0$ since $E_{C_{2x}\vec{k},\eta}= E_{\vec{k},\eta}$ implies that $\mc{T}_{xx}^{(2)}=0$.

To estimate the relative strengths of these two effects, we compute $\zeta(D)$ assuming a constant momentum-independent scattering rate, $\tau_{\vec{k},\eta}$\,$=$\,$\tau$, and using the eigenenergies $E_{\vec{k},\eta}$ obtained numerically from the continuum model with an unscreened displacement field $D$\,$=$\,$\Delta_0/L$. 
We find that as $\Delta_0$ is tuned from $-8$ to $+8$~meV, $\zeta$ varies between $\pm2.5^\circ$. Since this additional $D$-induced phase offset (by $\zeta$) of $\rho_T(2\Omega)$ in \equref{FinalFormOfRhoT} is significantly smaller than that due to the direct rotation of the director [$2\Delta \varphi \approx \pm12^\circ$ in the same range; see Fig.~\ref{fig:Rotation}(b)], transverse resistivity measurements provide a convenient method to probe the nematic director $\varphi$ directly, as well as its dependence on the electric field.

\section{Discussion and outlook} \label{sec:conclusions}
We investigated the nature of the nematic phase recently observed \cite{rubioverdu2020universal} experimentally in tDBG. A careful comparison between the STM data of \refcite{rubioverdu2020universal} and theoretical calculations using the continuum model for the spatial and energy dependencies of the LDOS in the nematic phase reveal that the most likely candidate to explain the experimental observations is a moir\'e nematic order parameter. In this case, the nematic instability is driven by emergent degrees of freedom associated with the moir\'e superlattice, rather than being a byproduct of the known nematic instability that takes place in bilayer graphene \cite{Mayorov2011}. This result suggests that the nematic phase is correlation-driven, and related to the physics of the moir\'e flat bands. Given that the same ingredients are present in several graphene-based moir\'e systems, this indicates that nematic order may be a more universal property of these types of systems, similar to the correlated insulating phases. 

We also showed that, quite generically, for a system with threefold rotational symmetry that displays Potts-nematic order, an external electric field can be used to rotate the electronic nematic director, unlocking it from the high-symmetry crystalline directions. This is in contrast to the more usual case of systems with fourfold rotational symmetry, where the director of the Ising-nematic order parameter is always fixed along high-symmetry directions, and only the magnitude of the order parameter can be tuned by external perturbations such as transverse strain or perpendicular magnetic fields \cite{Maharaj2017}. This result highlights the higher degree of tunability of Potts-nematic order as compared to Ising-nematic order, making this system a closer electronic analogue to the continuous nematic order parameter of classical liquid crystals. Interestingly, previous work on tBG \cite{2019arXiv191111367F} has shown that a spontaneous breaking of the $C_{2x}$ symmetry can take place in uniaxially strained devices that have an underlying nematic instability, depending on whether the external strain is compressive or tensile. 
Our current results show that a spontaneous electric field should emerge below this transition as well.

Finally, we demonstrated that this general coupling between the electric field and the nematic director is particularly sizable in tDBG, due to the sensitivity of its band structure to the displacement field. Specifically, we found that the nematic director can be rotated by about $10^{\circ}$ by sweeping the displacement field over an experimentally accessible range of values. We showed that this electric-field-induced rotation of the nematic director is manifested most prominently in measurements of the transverse resistivity with respect to a generic axis that does not coincide with the high-symmetry crystalline axes. If other correlated graphene-based moir\'e superlattice systems which are strongly susceptible to displacement fields (such as twisted trilayer graphene \cite{MirrSymTrilayer1,MirrSymTrilayer2} or tDBG with alternative stacking configurations \cite{koshino2019band}), also exhibit nematic order, our proposed electrical control mechanism and transport-based probes will be expected to apply as well.
Assuming the resistivity anisotropy is large in tDBG or other related moir\'e systems, 
our results open up the interesting possibility of designing devices that exploit the electric-field control of the electronic nematic director for future applications.

\acknowledgements
We thank D. Kennes, L. Klebl, H. Ochoa, and A. Rubio for fruitful discussions. RS acknowledges support from Subir Sachdev under the National Science Foundation Grant No. DMR-2002850. CRV acknowledges support from the European Union Horizon 2020 research and innovation program under the Marie Skłodowska-Curie grant agreement No. 844271. ANP and ST were supported by the Air Force Office of Scientific Research (AFOSR) via grant FA9550-16-1-0601. RMF was supported by the U. S. Department of Energy, Office of Science, Basic Energy Sciences, Materials Sciences and Engineering Division, under Award No. DE-SC0020045.

\appendix

\section{Continuum model for tDBG}
\label{app:model}

In this appendix, we briefly summarize the continuum model \cite{bistritzer2011moire} for tDBG introduced by \citet{koshino2019band} and tabulate the parameters employed for our calculations in the main text. The lattice vectors associated with each unrotated bilayer graphene (BLG) sheet are $\vec{a}_1$\,$=$\,$a(1,0)$ and $\vec{a}_2$\,$=$\,$a(1/2,\sqrt{3}/2)$, where $a$\,$\approx$\,$0.246\,\mathrm{nm}$ is the lattice constant of graphene. Correspondingly, in momentum space, the two reciprocal lattice vectors are $\vec{b}_1$\,$=$\,$(2\pi/a)(1,-1/\sqrt{3})$ and $\vec{b}_2 = (2\pi/a)(0,2/\sqrt{3})$. Upon the application of a relative twist between the pair of BLGs, the lattice vectors of the $l$-th BLG are modified to $\vec{a}_i^{(l)}$\,$=$\,$R(\mp \theta/2)\vec{a}_i$ with $\mp$ for $l$\,$=$\,$1,2$, $R (\theta)$ being the matrix for rotations by angle $\theta$; their reciprocal lattice counterparts are then given by $\vec{b}_i^{(l)}$\,$=$\,$R(\mp \theta/2)\,\vec{b}_i$. 
The reciprocal lattice vectors of the ensuing moir\'{e} pattern, in the limit of small $\theta$, are 
$ \vec{G}^{\rm M}_{i} = \vec{b}^{(1)}_i$\,$-$\,$\vec{b}^{(2)}_i \, (i = 1,2)$.
Note that the Dirac points of graphene are located at $\vec{K}^{(l)}_\eta = -\eta\, [2\vec{b}^{(l)}_1+\vec{b}^{(l)}_2]/3$ in valley $\eta=\pm 1$ for the $l$-th BLG.

Labeling the two sublattices  on layer $\ell=1,2,3,4$ by $A_{\ell},B_\ell$, the continuum Hamiltonian---at small twist angles $\theta\, (\ll 1)$---is expressed in the Bloch basis of carbon's $p_z$ orbitals, $(A_1,B_1,A_2, B_2, A_3, B_3, A_4, B_4)$ as
 \begin{align}
&	
{H}^{}_{\textrm{AB-AB}} = 
	\begin{pmatrix}
		H^{}_0(\vec{k}_1) & S^\dagger(\vec{k}_1) & & \\
		S(\vec{k}_1) & H'_0(\vec{k}_1) & U^\dagger  &\\
		& U & H^{}_0(\vec{k}_2) & S^\dagger(\vec{k}_2)  \\
		& & S(\vec{k}_2) & H'_0(\vec{k}_2) \\
	\end{pmatrix}  + V,
	\label{eq_AB-AB}
\end{align}
where $ \vec{k}_l = R(\pm \theta/2)({\vec{k}}-\vec{K}^{(l)}_\eta)$
with $\pm$ for $l$\,$=$\,$1,2$. Introducing the notation $k_\pm$\,$=$\,$\eta k_x \pm i k_y$, the various terms in ${H}_{\textrm{AB-AB}}$ are given by $2\times2$ matrices:
\begin{align}
\nonumber  H^{}_0(\vec{k}) 
&=
\begin{pmatrix}
0  & -\hbar v k_- \\
-\hbar v k_+ & d
\end{pmatrix},
\,
H'_0(\vec{k}) 
=
\begin{pmatrix}
d  & -\hbar v k_- \\
-\hbar v k_+ & 0
\end{pmatrix}, 
\\ 
S(\vec{k}) 
&=
\begin{pmatrix}
\hbar v_4 k_+  & \gamma_1 \\
\hbar v_3 k_-  & \hbar v_4 k_+
\end{pmatrix}.
\end{align}
Of these, $H_0$ and $H'_0$ above are the Hamiltonians of monolayer graphene \cite{moon2013optical,koshino2018maximally}, with a rescaled band velocity $v$ of $\hbar v /a$\,$=$\,$ 2.776\,$eV.
The parameter $d = 0.050\,$eV accounts for a small additional on-site potential at the so-called \textit{dimer} sites, where the $A_1$ atoms of the first layer are positioned directly above the $B_2$ atoms of the second \cite{mccann2013electronic}.
The matrix $S$ captures the interlayer coupling of AB-stacked BLG; $\gamma_1$\,$=$\,$0.4\,$eV denotes the coupling between the dimer sites, whereas $v_3$ and $v_4$ encode the diagonal hopping elements $\gamma_3$\,$=$\,$0.32\,$eV and $\gamma_4$\,$=$\,$0.044\,$eV as $v_i = (\sqrt{3}/2) \gamma_i a /\hbar \, (i=3,4)$ \cite{mccann2013electronic}, and represent trigonal warping \cite{mohan2020trigonal} of the energy band and electron-hole asymmetry, respectively, in AB-stacked BLG.

The couplings between the atoms in the second and third layers are encompassed by the moir\'{e} interlayer hopping matrix $U$ in Eq.~\eqref{eq_AB-AB} given by
\cite{bistritzer2011moire,moon2013optical,koshino2018maximally}
\begin{align}
 U &= 
\begin{pmatrix}
u & u'
\\
u' & u
\end{pmatrix}
+
\begin{pmatrix}
u & u'\omega^{-\eta}
\\
u'\omega^\eta & u
\end{pmatrix}
e^{i\, \eta \,\vec{G}^{\rm M}_1\cdot\vec{r}}
\nonumber\\
& 
 +
\begin{pmatrix}
u & u'\omega^\eta
\\
u'\omega^{-\eta} & u
\end{pmatrix}
e^{i \,\eta\, (\vec{G}^{\rm M}_1+\vec{G}^{\rm M}_2)\cdot\vec{r}}; \quad\omega =e^{2\pi i/3},
\label{eq_interlayer_matrix}
\end{align}
with $u$\,$=$\,$0.0797$\,eV and $u'$\,$=$\,$0.0975$\,eV \cite{koshino2018maximally}. It is noteworthy, however, that $u$\,$\ne$\,$u'$ as this difference between the diagonal ($u$) and off-diagonal ($u'$) amplitudes describes the crucial lattice relaxation effects \cite{koshino2019band,koshino2018maximally,nam2017lattice,tarnopolsky2019origin,lucignano2019crucial,PhysRevResearch.2.043127}. In reciprocal space, the coupling $U$ hybridizes the eigenstates at a Bloch vector $\vec{k}$ in the moir\'{e} Brillouin zone with the states at $\vec{q}$\,$=$\,$\vec{k}$\,$+$\,$\vec{G}$, where $\vec{G} = m_1  \vec{G}^{\rm M}_1 + m_2  \vec{G}^{\rm M}_2$ for $m_1, m_2 \in \mathbb{Z}$.

As discussed in much detail in the main text, an important tuning knob in the system is the potential difference between the top and bottom graphene layers arising from the perpendicular displacement field applied in the experimental setup. Assuming that the electric field is constant, the interlayer asymmetric potential---$V$ in Eq.~\eqref{eq_AB-AB}---can be written as
\begin{alignat}{1}
V = \mathrm{diag} \left(\frac{3\,\Delta_0}{2}\, \mathds{1}, 
\,\, \frac{\Delta_0}{2}\,\mathds{1}, 
\,\,-\frac{\Delta_0}{2}\, \mathds{1}, 
\,\, -\frac{3\,\Delta_0}{2}\, \mathds{1} \right),
\end{alignat}
where $\Delta_0$ is the difference in electrostatic energy between adjacent layers, and $\mathds{1}$ indicates the $2\times 2$ unit matrix. This is indeed the parametrization for the model used in Sec.~\ref{sec:EField} to study the effects of the displacement field. 

For a more detailed comparison with the STM experiments, however, it is necessary to take electronic screening into consideration. The effect of an applied displacement field on the electronic structure of twisted double-bilayer graphene can be captured in the continuum model by requiring that the onsite energy for each monolayer Hamiltonian be a function of that layer's vertical position $z$ within the stack.  In the limit of uniform screening, the interlayer potential difference is given simply by $\Delta_0 = L D/\epsilon$, where $L$ is the $c$-axis lattice constant, $D$ is the applied displacement field, and $\epsilon$ is the dielectric constant of the double-bilayer stack which we take to be $\sim12$.  To first order, the effect of screening can be described by a simple rescaling of the interlayer potential through the parameter $\epsilon$. In reality, however, the displacement field will tend to redistribute charge between the layers, thereby modulating the strength of the interlayer potential in a nonlinear fashion.  We capture this in our model by self-consistently solving for the layer-dependent onsite energies in a manner that has previously proven successful for untwisted graphene multilayers  \cite{koshino2010interlayer}.  In practice, this is achieved by initializing the calculation with an assumed set of layer potentials $\Delta(z)$, and solving for the resulting charge density $n(z)$ on each layer.  This gives rise to a new set of layer potentials, $\Delta'(z)$, given by the sum of the external displacement field and the field due to the redistributed charge densities $n(z)$.  We iterate this procedure until each $\Delta'(z)$ matches the corresponding input $\Delta(z)$ to within $1$~meV and use this converged set of layer potentials for the calculations presented in the main text.
Hence, for including a self-consistently calculated screened electric field, as in Sec.~\ref{sec:LDOS}, it is more convenient to use the formulation
\begin{alignat}{1}
V = \mathrm{diag} \left(\Delta_1\, \mathds{1}, 
\,\, \Delta_2\,\mathds{1}, 
\,\,\Delta_3\, \mathds{1}, 
\,\,\Delta_4\, \mathds{1} \right).
\end{alignat}
In this work, at a filling fraction of $\nu $\,$=$\,$0.475$, we take the on-site potentials to be $\Delta_1$\,$=$\,$4.079$\,meV, $\Delta_2$\,$=$\,$1.021$\,meV, $\Delta_3$\,$=$\,$-1.537$\,meV, and $\Delta_4$\,$=$\,$-3.563$\,meV for layers 1 through 4, respectively. For a generic value of $\nu$, however, $\Delta_i$\,($i=1,\ldots,4)$ would assume different values.

Having established the structure of the the Hamiltonian~\eqref{eq_AB-AB}, we carry out a numerical diagonalization  in momentum space (keeping only finitely many wavevectors $\vec{q}$ within the cutoff circle $\lvert \vec{q}- (\vec{K}^{(1)}_\eta+ \vec{K}^{(2)}_\eta)/2\rvert < 4 |\vec{G}^{\rm M}_1|$) to obtain the low-energy wavefunctions and eigenenergies \cite{PhysRevResearch.1.013001}.  
This calculation can be performed independently for each valley because the intervalley coupling is negligible at small twist angles.  The band structure obtained from such a computation is presented in Fig.~\ref{fig:Lattice}(b).

\bibliography{Refs}
\end{document}